\documentclass[iop,floatfix]{emulateapj}

\RequirePackage{graphicx}
\RequirePackage{color}
\RequirePackage{code}
\input{astro.sty}

\usepackage[T1]{fontenc}

\def\plotext{pdf}

\shorttitle{Pixel Analysis in PS1}
\shortauthors{E.A. Magnier et al}
\begin{document}
\title{Pan-STARRS Pixel Analysis : Source Detection and Characterization}

\def\IfA{1}
\def\Princeton{2}
\def\DUR{3}
\def\CfA{2}

\author{
Eugene A. Magnier,\altaffilmark{\IfA}
W.~E. Sweeney,\altaffilmark{\IfA}
K.~C. Chambers,\altaffilmark{\IfA} 
H.~A. Flewelling,\altaffilmark{\IfA}
M. E. Huber,\altaffilmark{\IfA}
P.~A. Price,\altaffilmark{\Princeton}
C. Z. Waters,\altaffilmark{\IfA}
L. Denneau,\altaffilmark{\IfA}
P. Draper,\altaffilmark{\DUR}
R. Jedicke,\altaffilmark{\IfA}
K. W. Hodapp,\altaffilmark{\IfA}
N. Kaiser,\altaffilmark{\IfA}
R.-P. Kudritzki,\altaffilmark{\IfA}
N. Metcalfe,\altaffilmark{\DUR}
C.~W. Stubbs,\altaffilmark{\CfA}
R. J. Wainscoat\altaffilmark{\IfA}
} 

\altaffiltext{\IfA}{Institute for Astronomy, University of Hawaii, 2680 Woodlawn Drive, Honolulu HI 96822}
\altaffiltext{\Princeton}{Department of Astrophysical Sciences, Princeton University, Princeton, NJ 08544, USA}
\altaffiltext{\DUR}{Department of Physics, Durham University, South Road, Durham DH1 3LE, UK}
\altaffiltext{\CfA}{Harvard-Smithsonian Center for Astrophysics, 60 Garden Street, Cambridge, MA 02138}
\begin{abstract}

Over 3 billion astronomical sources have been detected in the more
than 22 million orthogonal transfer CCD images obtained as part of the
Pan-STARRS\,1 $3\pi$ survey.  Over 85 billion instances of those
sources have been automatically detected and characterized by the
Pan-STARRS Image Processing Pipeline photometry software,
\ippprog{psphot}.  This fast, automatic, and reliable software was
developed for the Pan-STARRS project, but is easily adaptable to
images from other telescopes.  We describe the analysis of the
astronomical sources by \ippprog{psphot} in general as well as for the
specific case of the 3rd processing version used for the first public
release of the Pan-STARRS $3\pi$ survey data.
\end{abstract}


\section{Introduction}
\label{sec:intro}


The 1.8m Pan-STARRS\,1 telescope is located on the summit of Haleakala
on the Hawaiian island of Maui.  The wide-field optical design of the
telescope \citep{2004SPIE.5489..667H} produces a 3.3 degree field of view with
low distortion and minimal vignetting even at the edges of the
illuminated region.  The optics and natural seeing combine to yield
good image quality: 75\% of the images have full-width half-max values
less than (1.51, 1.39, 1.34, 1.27, 1.21) arcseconds for (\grizy), with
a floor of $\sim 0.7$ arcseconds.

The \PSONE\ camera \citep{2009amos.confE..40T}, known as GPC1, consists of a
mosaic of 60 back-illuminated CCDs manufactured by Lincoln Laboratory.
The CCDs each consist of an $8\times8$ grid of $590 \times 598$
pixel readout regions, yielding an effective $4846 \times 4868$
detector.  Initial performance assessments are presented in
\cite{2008SPIE.7014E..0DO}.  Routine observations are conducted remotely from the
Advanced Technology Research Center in Kula, the main facility of the
University of Hawaii's Institute for Astronomy (IfA) operations on Maui.
The Pan-STARRS1 filters and photometric system have already been
described in detail in \cite{2012ApJ...750...99T}.

For nearly 4 years, from 2010 May through 2014 March, this telescope
was used to perform a collection of astronomical surveys under the
aegis of the Pan-STARRS Science Consortium.  The majority of the time
(56\%) was spent on surveying the $\frac{3}{4}$ of the sky north of
$-30$ Declination with \grizy\ filters in the so-called $3\pi$ Survey.
Another $\sim 25\%$ of the time was concentrated on repeated deep
observations of 10 specific fields in the Medium-Deep Survey.  The
rest of the time was used for several other surveys, including a
search for potentially hazardous asteroids in our solar system.  The
details of the telescope, surveys, and resulting science publications
are described by \cite{chambers2017}.

Pan-STARRS produced its first large-scale public data release, Data
Release 1 (DR1) on 16 December 2016.  DR1 contains the results of the
third full reduction of the Pan-STARRS $3\pi$ Survey archival data,
identified as PV3.  Previous reductions \citep[PV0, PV1, PV2;
  see][]{magnier2017.datasystem} were used internally for pipeline
optimization and the development of the initial photometric and
astrometric reference catalog \citep{magnier2017.calibration}.  The
products from these reductions were not publicly released, but have
been used to produce a wide range of scientific papers from the
Pan-STARRS 1 Science Consortium members \citep{chambers2017}.  DR1
contained only average information resulting from the many individual
images obtained by the $3\pi$ Survey observations.  A second data
release, DR2, was made available 28 January 2019.  DR2 provides
measurements from all of the individual exposures, and include an
improved calibration of the PV3 processing of that dataset.

This is the fourth in a series of seven papers describing the
Pan-STARRS1 Surveys, the data reduction techniques and the resulting
data products.  This paper (Paper IV) describes the details of the
source detection and photometry, including point-spread-function and
extended source model fitting, and the techniques for ``forced''
photometry measurements.  The software describe here was used with a
single consistent set of parameters for the complete PV3 analysis,
used for both DR1 and DR2.

\citet[][Paper I]{chambers2017}
provide an overview of the Pan-STARRS System, the design and
execution of the Surveys, the resulting image and catalog data
products, a discussion of the overall data quality and basic
characteristics, and a brief summary of important results.

\citet[][Paper II]{magnier2017.datasystem}
describe how the various data processing stages are organized and implemented
in the Imaging Processing Pipeline (IPP), including details of the 
the processing database which is a critical element in the IPP infrastructure . 

\citet[][Paper III]{waters2017} describe the details of the pixel
processing algorithms, including detrending, warping, and adding (to
create stacked images) and subtracting (to create difference images)
and resulting image products and their properties.


\citet[][Paper V]{magnier2017.calibration}
describe the final calibration process, and the resulting photometric and astrometric quality.

\citet[][Paper VI]{flewelling2017}
describe  the details of the resulting catalog data and its organization in the Pan-STARRS database. 

\citet[][Paper VII]{huber2017} describe the Medium Deep Survey in
detail, including the unique issues and data products specific to that
survey. The Medium Deep Survey is not part of Data Releases 1 or 2 and
will be made available in a future data release.


\section{Background}

The photometric and astrometric precision goals for the Pan-STARRS\,1
surveys were quite stringent: photometric accuracy of 10
millimagnitudes, relative astrometric accuracy of 10 milliarcseconds
and absolute astrometric accuracy of 100 milliarcseconds with respect
to the ICRS reference stars.

An additional constraint on the Pan-STARRS analysis system comes from
the high data rate.  PS1 produces typically $\sim 500$ exposures per
night, corresponding to $\sim 750$ billion pixels of imaging data.
The images range from high galactic latitudes to the Galactic bulge,
so large numbers of measurable stars can be expected in much of the
data.  The combination of the high precision goals of the astrometric
and photometric measurements and the high data rate (and a finite
computing budget) mean that the process of detecting, classifying, and
measuring the astronomical sources in the image data stream in a
timely fashion are a significant challenge.

In order to achieve these ambitious goals, the source detection,
classification, and measurement process must be both precise and
efficient.  Not only is it necessary to make a careful measurement of
the flux of individual sources, it is also critical to characterize
the image point-spread-function, and its variations across the field
and from image to image.  Since comparisons between images must be
reliable, the measurements must be stable for both photometry and
astrometry.

A variety of astronomical software packages perform the basic source
detection, measurement, and classification tasks needed by the
Pan-STARRS IPP.  Each of these programs have their own advantages and
disadvantages.  Below we discuss some of the most widely used of these
other packages, highlighting the features of the programs which are
particularly desirable, and noting aspects of the programs which are
problematic for the IPP.

\begin{itemize}

\item DoPhot : analytical fitted model with aperture corrections.
  pro: well-tested, stable code.  con: limited range of models,
  algorithm converges slowly to a PSF model, limited tests of PSF
  validity, inflexible code base, fortran \citep{1993PASP..105.1342S}.

\item DAOPhot : Pixel-map PSF model with analytical component.  pro:
  well-tested, high-quality photometry.  con: Difficult to use in an
  automated fashion, does it handle 2D variations well? \citep{1987PASP...99..191S}.

\item Sextractor : pure aperture measurement with rudimentary source
  subtraction.  pro: fast, widely used, easy to automate.  con: poor
  source separation in crowded regions, PSF-modeling was only in beta,
  not widely used at the time \citep{sextractor}.

\item galfit : detailed galaxy modeling.  not a multi-source PSF
  analysis tool.  con: does not provide a PSF model, not easily
  automated.  very detailed results in very slow processing.  only a
  galaxy analysis program \citep{2002AJ....124..266P}.

\item SDSS phot : con: tightly integrated into the SDSS software
  environment \citep{2001ASPC..238..269L}.

\end{itemize}

When the IPP development was starting, the existing photometry
packages either did not meet the accuracy requirements or required too
much human intervention to be considered for the needs of PS1.  In the
case of the SDSS Photo tool, the software was judged to be too tightly
integrated to the architecture of SDSS to be easily re-integrated into
the Pan-STARRS pipeline.  A new photometry analysis package was
developed using lessons learned from the existing photometry systems.
In the process, the source analysis software was written using the
data analysis C-code library written for the IPP, \code{psLib}
\citep{magnier2017.datasystem}.  Components of the photometry code
were integrated into the IPP's mid-level astronomy data analysis
toolkit called \code{psModules} \citep{magnier2017.datasystem}.  The
resulting software, `\ippprog{psphot}', can be used either as a
stand-alone C program, or as a set of library functions which may be
integrated into other programs

Several variants of \ippprog{psphot} have been used in the PS1 PV3
analysis.  The main variant of \ippprog{psphot} operates on a single
image, or a group of related images representing the data read from a
camera in a single exposure.  The images are expected to have already
been detrended so that pixel values are linearly related to the flux.
The gain may be specified by the configuration system, or a variance
image may be supplied.  A mask may also be supplied to mark good, bad,
and suspect pixels.  This variant of \ippprog{psphot} can be called as a
stand-alone program, also called \ippprog{psphot}.  In standard IPP
operations, this variant is used as a library call within the analysis
program \ippprog{ppImage} during the \ippstage{chip} analysis stage.

The variant called \ippprog{psphotStack} accepts a set of images, each
representing the same patch of sky in a different filter, nominally
the full $grizy$ filter set for the analysis of the PS1 PV3 stack
images, though where insufficient data were available in a given
filter, a subset of these filters was processed as a group.  As
discussed in detail below, the \ippprog{psphotStack} analysis includes the
capability of measuring forced PSF photometry in some filter images
based on the position of sources detected in the other filters.  It
also include an option to convolve the set of images to a single,
common PSF size across the filters for the purpose of fixed aperture
photometry.

Another variant of \ippprog{psphot} used in the PV3 analysis is called
\ippprog{psphotFullForce}.  In this variant, a set of image all representing the
same pixels are processed together, with the positions of sources to
be analyzed loaded from a supplied file.  In this variant of the
analysis, sources are not discovered -- only the supplied sources are
considered.  PSF models are determined for each exposure and the
forced PSF photometry is measured for all sources.  A subset of
sources may also be used to measure forced galaxy shape parameters.
As described below, a grid of galaxy models are fitted based on the
supplied guess model.  

\section{\nocode{psphot} Design Goals}


\ippprog{psphot} has a number of important requirements that it must
meet, and a number of design goals which we believe will help to make
it usable in a wide range of circumstances.  The critical
astronomy-driven measurement goals of the Pan-STARRS project, which
drive the design of \ippprog{psphot}, are the photometric accuracy
goal (10 millimagntudes) and the astrometric accuracy goal (10
milliarcseconds).  For \ippprog{psphot}, the photometry accuracy goal
implies that the measured photometry of stellar sources must be
substantially better than this 10 mmag goal since the photometry error
per image is combined with an error in the flat-field calibration and
an error in measuring the atmospheric effects.  We have set a goal for
\ippprog{psphot} of 3mmag photometric consistency for bright stars
between pairs of images obtained in photometric conditions at the same
pointing, ie to remove sensitivity to flat-field errors.  This goal
splits the difference between the three main contributors and still
allows some leeway.  This requirement must be met for well-sampled
images and images with only modest undersampling.

The relative astrometric calibration depends on the consistency of the
individual measurements.  The measurements from \ippprog{psphot} must
be sufficiently representative of the true source position to enable
astrometric calibration at the 10mas level.  The error in the
individual measurements will be folded together with the errors
introduced by the optical system, the effects of seeing, and by the
available reference catalogs.  We have set a goal for \ippprog{psphot}
of 5mas consistency between the true source postion and the measured
position given reasonable PSF variations under simulations.  This
level must be reached for images with 250 mas pixels, implying
\ippprog{psphot} must introduce measurement errors less than 1/50th of
a pixel. The choice of 32 bit floating point data values for the
source centroids places a numerical limit of 1e-7 on the accuracy of a
pixel relative to the size of a chip (since a single data value is
used for X or Y).  For the $4800^2$ GPC chips, this yields a limit of
about 0.25 milliarcsecond.


The design goals for \ippprog{psphot} are chosen to make the program flexible,
general, and able to meet the unknown usage cases future projects may
require:

\begin{itemize}
\item {\bf Flexible PSF model} Different image sources require
  different ways of representing the PSF.  Ideally, both analytical
  and pixel-based versions should be possible.

\item {\bf PSF spatial variation} Most images result in some spatial
  PSF variations at a certain level.  The PSF representation should
  naturally incorporate 2-D variations.

\item {\bf Flexible non-PSF models} \ippprog{psphot} must be able to represent
  PSF-like sources as well as non-PSF sources (e.g., galaxies).  It
  must be easy to add new source models as interesting representations
  of sources are invented.

\item {\bf Clean code base} \ippprog{psphot} should incorporate a high-degree of
  abstraction and encapsulation so that changes to the code structure
  can be performed without pulling the code apart and starting from scratch.

\item {\bf PSF validity tests} \ippprog{psphot} should include the ability to
  choose different types of PSF models for different situations, or to
  provide the user with methods for assessing the different PSF models.

\item {\bf Careful systematic corrections} \ippprog{psphot} must carefully
  measure and correct for the photometric and astrometric trends
  introduced by using analytical PSF models.

\item {\bf User Configurable} \ippprog{psphot} should allow users to change the
  options easily and to allow different approaches to the analysis.

\end{itemize}

\section{\nocode{psphot} Analysis Process}

\subsection{Overview}

The \ippprog{psphot} analysis is divided into several major stages, as
listed below.  

\begin{enumerate}
\item {\bf Image Preparation} Load data, characterize the image
  background, load or construct variance and mask images.

\item {\bf Initial Source Detection} Smooth, find peaks, measure basic
  properties.

\item {\bf PSF Determination} Select PSF candidates, perform model
  fits, build PSF model from fits, select best PSF model class.

\item {\bf Bright Source Analysis} Fit sources with PSFs, determine
  PSF validity, subtract PSF-like sources, fit non-PSF model(s),
  select best model class, subtract model.

\item {\bf Faint Source Analysis} Detect low-level sources, measure
  properties (aperture or PSF)

\item {\bf Extended Source Analysis} Detailed measurements relevant to
  galaxies and/or other extended (non-PSF) sources.

\item {\bf Aperture corrections} Measure the curve-of-growth, spatial
  aperture variations, and background-error corrections.  

\item {\bf Output} Write out sources in selected format, write out
  difference image, variance image, etc, as selected.
\end{enumerate}

Table~\ref{tab:measurements} lists the types of
analyses performed by \ippprog{psphot}, specifying which of the
\ippprog{psphot} usage cases performs the given analysis.  The table
also provides a reference to the section of this paper in which the
analysis is described.  Not all analyses are relevant to all sources
in all images.  The table identifies thoses cases where the analyses
are applied to only a subset of all sources.  

\ippprog{psphot} is highly configurable.  Users may choose via the configuration
system which of the above analyses are performed.  This is useful for
testing, but also allows for specialized use cases.  For example, the
PSF model may already be available from external information, in which
case the PSF modeling stage can be skipped.

\begin{table*}
\caption{\label{tab:measurements} \nocode{psphot} measurements performed} 
\begin{center}
\footnotesize
\begin{tabular}{lccccll}
\hline
\hline
{\bf Measurement} & {\bf Camera} & {\bf Stack} & {\bf Forced Warp} & {\bf Diff} & {\bf Section} & {\bf Which} \\
\hline
  Background Subtraction     & Y & Y & Y & N$^1$ & \ref{sec:image.preparation}      & N/A \\
  Peaks                      & Y & Y & N & Y     & \ref{sec:peaks}                  & All \\
  Footprints                 & Y & Y & N & Y     & \ref{sec:footprints}             & All \\
  Moments                    & Y & Y & Y & Y     & \ref{sec:moments}                & All \\
  PSF Model                  & Y & Y & Y & N$^2$ & \ref{sec:PSF.Model}              & Uses bright, unsat. stars \\
  Bright Star Profile        & Y & Y & N & Y     & \ref{sec:very.bright.star}       & Saturated Stars \\
  Radial Profiles v1         & Y & Y & N & Y     & \ref{sec:radial.profile}         & All \\
  Kron Fluxes                & Y & Y & Y & Y     & \ref{sec:kron.mags}              & All \\
  Source-Size Tests          & Y & Y & N & Y     & \ref{sec:source.size}            & All \\
  Non-Linear PSF Fits        & Y & Y & N & N     & \ref{sec:nonlinear.psf.model}    & $S/N > 20$ \\
  Unconvolved Galaxy Model   & Y & Y & N & N     & \ref{sec:nonlinear.galaxy.model} & $S/N > 20$, extended \\
  Unconvolved Streak Model   & N & N & N & Y     & \ref{sec:nonlinear.galaxy.model} & $S/N > 20$, extended \\
  Linear PSF Fits            & Y & Y & Y & Y     & \ref{sec:faint.psf.model}        & All \\
  Radial Profiles v2         & Y & Y & N & Y     & \ref{sec:radial.profile.v2}      & Gal. Latitude Cut \\
  Petrosian Fluxes           & N & Y & Y & N     & \ref{sec:petrosian}              & Gal. Latitude Cut \\
  Convolved Galaxy Models    & N & Y & N & N     & \ref{sec:galaxy.conv.fit}        & Gal. Latitude Cut, mag cut \\
  Fixed Aperture Photometry  & N & Y & Y & N     & \ref{sec:fixed.aperture.photom}  & All \\
  Convolved, Fixed Apertures & N & Y & N & N     & \ref{sec:fixed.aperture.photom}  & All \\
  Aperture Corrections       & Y & Y & Y & N     & \ref{sec:aperture.correction}    & All \\
  Forced PSF Fluxes          & N & N & Y & N     & \ref{sec:psf.forced.fit}         & All \\
  Forced Galaxy Models       & N & N & Y & N     & \ref{sec:galaxy.forced.fit}      & Have Stack Galaxy Models \\
  Lensing Parameters         & N & Y & Y & N     &                                  & All \\
\hline
\multicolumn{5}{l}{$^1$ Background subtraction is performed by {\tt ppSub} before calling {\tt psphot}} \\
\multicolumn{5}{l}{$^2$ PSF modeling is perform by {\tt ppSub} on the input warps before calling {\tt psphot}} \\
\end{tabular}
\end{center}
\end{table*}

\subsection{Informational and Warning Bit Flags}

During the \ippprog{psphot} analysis, there are a wide variety of
conditions which are identified by the analysis software.  As part of
the output data for each detected source, two fields are provided
which encode these conditions as bit values in the two 32-bin
integers.  The following two tables list the individual bit values in
these two fields.  These informational and warning bits are described
in more detail later in this article.
Table~\ref{tab:det_flag_values} lists the flags recorded in the output
field \ippmisc{FLAGS}.  When data from \ippprog{psphot} is loaded into
a DVO database \citep{magnier2017.calibration}, these values are
stored in the field \code{Measure.photFlags} and exposed in the public
database \citep[PSPS][]{flewelling2017} in the fields
\code{Detection.infoFlag}, \code{StackObjectThin.XinfoFlag} (where
\code{X} is one of {$grizy$}), and
\code{ForcedWarpMeasurement.FinfoFlag}.
Table~\ref{tab:det_flag2_values} lists the flags recorded in the
output field \ippmisc{FLAGS2}.  When data from \ippprog{psphot} is
loaded into a DVO database \citep{magnier2017.calibration}, these
values are not currently loaded, but they are exposed in PSPS in the fields
\code{Detection.infoFlag2}, \code{StackObjectThin.XinfoFlag2} (where
\code{X} is one of {$grizy$}), and
\code{ForcedWarpMeasurement.FinfoFlag2}.

\begin{table*}
\caption{\label{tab:det_flag_values} \nocode{psphot} Detection Flag Values \#1} 
\begin{center}
\footnotesize
\begin{tabular}{lrl}
\hline
\hline
{\bf Flag Name} & {\bf Flag Value} & {\bf Description} \\
\hline
 PM\_SOURCE\_MODE\_PSFMODEL            & 0x00000001 & Source fitted with a psf model (linear or non-linear) \\
 PM\_SOURCE\_MODE\_EXTMODEL            & 0x00000002 & Source fitted with an extended-source model \\
 PM\_SOURCE\_MODE\_FITTED              & 0x00000004 & Source fitted with non-linear model (PSF or EXT; good or bad) \\
 PM\_SOURCE\_MODE\_FAIL                & 0x00000008 & Fit (non-linear) failed (non-converge, off-edge, run to zero) \\
 PM\_SOURCE\_MODE\_POOR                & 0x00000010 & Fit succeeds, but low-SN, high-Chisq, or large (for PSF -- drop?) \\
 PM\_SOURCE\_MODE\_PAIR                & 0x00000020 & Source fitted with a double psf \\
 PM\_SOURCE\_MODE\_PSFSTAR             & 0x00000040 & Source used to define PSF model \\
 PM\_SOURCE\_MODE\_SATSTAR             & 0x00000080 & Source model peak is above saturation \\
 PM\_SOURCE\_MODE\_BLEND               & 0x00000100 & Source is a blend with other sources$^1$ \\
 PM\_SOURCE\_MODE\_EXTERNAL            & 0x00000200 & Source based on supplied input position \\
 PM\_SOURCE\_MODE\_BADPSF              & 0x00000400 & Failed to get good estimate of object's PSF \\
 PM\_SOURCE\_MODE\_DEFECT              & 0x00000800 & Source is thought to be a defect \\
 PM\_SOURCE\_MODE\_SATURATED           & 0x00001000 & Source is thought to be saturated pixels (bleed trail) \\
 PM\_SOURCE\_MODE\_CR\_LIMIT           & 0x00002000 & Source has crNsigma above limit \\
 PM\_SOURCE\_MODE\_EXT\_LIMIT          & 0x00004000 & Source has extNsigma above limit \\
 PM\_SOURCE\_MODE\_MOMENTS\_FAILURE    & 0x00008000 & could not measure the moments \\
 PM\_SOURCE\_MODE\_SKY\_FAILURE        & 0x00010000 & could not measure the local sky \\
 PM\_SOURCE\_MODE\_SKYVAR\_FAILURE     & 0x00020000 & could not measure the local sky variance \\
 PM\_SOURCE\_MODE\_BELOW\_MOMENTS\_SN  & 0x00040000 & moments not measured due to low S/N.$^1$ \\
 PM\_SOURCE\_MODE\_BIG\_RADIUS         & 0x00100000 & poor moments for small radius, try large radius \\
 PM\_SOURCE\_MODE\_AP\_MAGS            & 0x00200000 & source has an aperture magnitude \\
 PM\_SOURCE\_MODE\_BLEND\_FIT          & 0x00400000 & source was fitted as a blend \\
 PM\_SOURCE\_MODE\_EXTENDED\_FIT       & 0x00800000 & full extended fit was used \\
 PM\_SOURCE\_MODE\_EXTENDED\_STATS     & 0x01000000 & extended aperture stats calculated \\
 PM\_SOURCE\_MODE\_LINEAR\_FIT         & 0x02000000 & source fitted with the linear fit \\
 PM\_SOURCE\_MODE\_NONLINEAR\_FIT      & 0x04000000 & source fitted with the non-linear fit \\
 PM\_SOURCE\_MODE\_RADIAL\_FLUX        & 0x08000000 & radial flux measurements calculated \\
 PM\_SOURCE\_MODE\_SIZE\_SKIPPED       & 0x10000000 & size could not be determined$^1$ \\
 PM\_SOURCE\_MODE\_ON\_SPIKE           & 0x20000000 & peak lands on diffraction spike \\
 PM\_SOURCE\_MODE\_ON\_GHOST           & 0x40000000 & peak lands on ghost or glint \\
 PM\_SOURCE\_MODE\_OFF\_CHIP           & 0x80000000 & peak lands off edge of chip \\
\hline
\multicolumn{3}{l}{$^1$ Not used for DR1 or DR2.} \\
\hline
\end{tabular}
\end{center}
\end{table*}

\begin{table*}
\caption{\label{tab:det_flag2_values} \nocode{psphot} Detection Flag Values \#2} 
\begin{center}
\footnotesize
\begin{tabular}{lrl}
\hline
\hline
{\bf Flag Name} & {\bf Flag Value} & {\bf Description} \\
\hline
 PM\_SOURCE\_MODE2\_DIFF\_WITH\_SINGLE      & 0x00000001 & diff source matched to a single positive detection \\
 PM\_SOURCE\_MODE2\_DIFF\_WITH\_DOUBLE      & 0x00000002 & diff source matched to positive detections in both images \\
 PM\_SOURCE\_MODE2\_MATCHED          	    & 0x00000004 & source generated based on another image \\
 PM\_SOURCE\_MODE2\_ON\_SPIKE               & 0x00000008 & $> 25\%$ of (PSF-weighted) pixels land on diffraction spike \\
 PM\_SOURCE\_MODE2\_ON\_STARCORE            & 0x00000010 & $> 25\%$ of (PSF-weighted) pixels land on starcore \\
 PM\_SOURCE\_MODE2\_ON\_BURNTOOL            & 0x00000020 & $> 25\%$ of (PSF-weighted) pixels land on burntool \\
 PM\_SOURCE\_MODE2\_ON\_CONVPOOR            & 0x00000040 & $> 25\%$ of (PSF-weighted) pixels land on convpoor \\
 PM\_SOURCE\_MODE2\_PASS1\_SRC              & 0x00000080 & source detected in first pass analysis \\
 PM\_SOURCE\_MODE2\_HAS\_BRIGHTER\_NEIGHBOR & 0x00000100 & peak is not the brightest in its footprint \\
 PM\_SOURCE\_MODE2\_BRIGHT\_NEIGHBOR\_1     & 0x00000200 & $flux_{\rm n} / (r^2 flux_{\rm p}) > 1$ \\
 PM\_SOURCE\_MODE2\_BRIGHT\_NEIGHBOR\_10    & 0x00000400 & $flux_{\rm n} / (r^2 flux_{\rm p}) > 10$ \\
 PM\_SOURCE\_MODE2\_DIFF\_SELF\_MATCH       & 0x00000800 & positive detection match is probably this source \\
 PM\_SOURCE\_MODE2\_SATSTAR\_PROFILE        & 0x00001000 & saturated source is modeled with a radial profile \\
 PM\_SOURCE\_MODE2\_ECONTOUR\_FEW\_PTS      & 0x00002000 & too few points to measure the elliptical contour \\
 PM\_SOURCE\_MODE2\_RADBIN\_NAN\_CENTER     & 0x00004000 & radial bins failed with too many NaN center bin \\
 PM\_SOURCE\_MODE2\_PETRO\_NAN\_CENTER      & 0x00008000 & petrosian radial bins failed with too many NaN center bin$^1$ \\
 PM\_SOURCE\_MODE2\_PETRO\_NO\_PROFILE      & 0x00010000 & petrosian not build because radial bins missing \\
 PM\_SOURCE\_MODE2\_PETRO\_INSIG\_RATIO     & 0x00020000 & insignificant measurement of petrosian ratio \\
 PM\_SOURCE\_MODE2\_PETRO\_RATIO\_ZEROBIN   & 0x00040000 & petrosian ratio in the 0th bin (likely bad) \\
 PM\_SOURCE\_MODE2\_EXT\_FITS\_RUN          & 0x00080000 & we attempted to run extended fits on this source \\
 PM\_SOURCE\_MODE2\_EXT\_FITS\_FAIL         & 0x00100000 & at least one of the model fits failed \\
 PM\_SOURCE\_MODE2\_EXT\_FITS\_RETRY        & 0x00200000 & trailed asteroid model fit was re-tried with new window \\
 PM\_SOURCE\_MODE2\_EXT\_FITS\_NONE         & 0x00400000 & ALL of the model fits failed \\
\hline
\multicolumn{3}{l}{$^1$ Not used for DR1 or DR2.} \\
\hline
\end{tabular}
\end{center}
\end{table*}

\subsection{Image Preparation}
\label{sec:image.preparation}

The first step is to prepare the image for detection of the
astronomical sources.  We need three separate images: the measured
flux (signal image), the corresponding variance image, and a mask
defining which pixels are valid and which should be ignored.  The
signal and variance images are represented internally as 32-bit
floating point values.  The variance and mask images may either
be provided by the user, or they may be automatically generated from
the input image, based on configuration-defined values for the image
gain, read-noise, saturation, and so forth.  For the function-call
form of the program, the flux image is provided in the API, and
references to the mask and variance are provided in the configuration
information.  As in the stand-alone C-program, the variance and mask may
be constructed automatically by \ippprog{psphot}.

The mask is represented as a 16-bit integer image in which a value of
0 represents a valid pixel.  Each of the 16 bits define different
reasons a pixel should be ignored.  This allows us to optionally
respect or ignore the mask depending on the circumstance.  For
example, in some cases, we ignore saturated pixels completely while in
other circumstances, it may be useful to know the flux value of the
saturated pixel.  In addition, the mask pixels are used to define the
pixels available during a model fit; those which should be ignored for
that specific fit are `marked' by setting a special bit (\code{MARK = 0x8000}).
The initial mask, if not supplied by the user or library calls, is
constructed by default from the image by applying three rules: 1)
Pixels which are above a specified saturation level are marked as
saturated.  The level is specified by the camera format keyword
\code{CELL.SATURATION}, which may specify a value or define a header
keyword which in turn specifies the value in the image header.  In the
case of PS1 PV3, the header keyword \code{MAXLIN} specifies the
saturation level for each chip \citep[see][]{waters2017}. 2) Pixels
which are below a user-defined value are considered unresponsive and
masked as dead.  (camera format keyword \code{CELL.BAD} = 0 for PS1
PV3).  3) Pixels which lie outside of a user-defined coordinate window
are considered non-data pixels (\eg, overscan) and are marked as
invalid.  (\ippprog{psphot} recipe keywords \code{XMIN}, \code{XMAX},
\code{YMIN}, \code{YMAX}, all set to 0 for PS1 PV3 -- invalid pixels
were specified for PS1 PV3 with a supplied mask image
\citep[see][]{waters2017}.

The library functions used by \ippprog{psphot} understand two types of
masked pixels: ``bad'' and ``suspect''.  Bad pixels are those which
should not be used in any operations, while suspect pixels are those
for which the reported signal may be contaminated or biased, but may
be usable in some contexts.  For example, a pixel with poor charge
transfer efficiency is likely to be too untrustworthy to use in any
circumstance, while a pixel in which persistence ghosts have been
subtracted might be useful for detection or even analysis of brighter
sources.  Table~\ref{tab:mask_values} lists the 16 bit values used for
PS1 mask images, along with their description \citep[see][for
  additional information]{waters2017}.

\begin{table*}
\caption{\label{tab:mask_values} \nocode{psphot} / GPC1 Mask Image Pixel Values} 
\begin{center}
\footnotesize
\begin{tabular}{lcccl}
\hline
\hline
{\bf Mask Name} & {\bf Mask Value} & {\bf Dynamic?} & {\bf Suspect?} & {\bf Description} \\
\hline
  DETECTOR & 0x0001 & N & N & A detector defect is present. \\
  FLAT     & 0x0002 & N & N & The flat field model does not calibrate the pixel reliably. \\
  DARK     & 0x0004 & N & N & The dark model does not calibrate the pixel reliably. \\
  BLANK    & 0x0008 & N & N & The pixel does not contain valid data. \\
  CTE      & 0x0010 & N & N & The pixel has poor charge transfer efficiency. \\
  SAT      & 0x0020 & Y & N & The pixel is saturated. \\
  LOW      & 0x0040 & Y & N & The pixel has a lower value than expected. \\
  SUSPECT  & 0x0080 & Y & Y & The pixel is suspected of being bad$^1$. \\
  BURNTOOL & 0x0080 & Y & Y & The pixel contain an burntool repaired streak. \\
  CR       & 0x0100 & Y & N & A cosmic ray is present. \\
  SPIKE    & 0x0200 & Y & Y & A diffraction spike is present. \\
  GHOST    & 0x0400 & Y & Y & An optical ghost is present. \\
  STREAK   & 0x0800 & Y & Y & A streak is present. \\
  STARCORE & 0x1000 & Y & Y & A bright star core is present. \\
  CONV.BAD & 0x2000 & Y & N & The pixel is bad after convolution with a bad pixel. \\
  CONV.POOR& 0x4000 & Y & Y & The pixel is poor after convolution with a bad pixel. \\
  MARK     & 0x8000 & X & X & An internal flag for temporarily marking a pixel. \\
\hline
\multicolumn{5}{l}{$^1$ The SUSPECT bit is generic and only
  used if a specific reason cannot be identified.}\\
\multicolumn{5}{l}{~~~It is overloaded on the same bit as BURNTOOL.}\\
\end{tabular}
\end{center}
\end{table*}

The variance image, if not supplied, is constructed by default from
the flux image using the configuration supplied gain and read noise
values to calculate the appropriate Poisson statistics for each pixel.
The parameters are determined based on the camera format keywords
\code{CELL.GAIN} and \code{CELL.READNOISE}, which in the case of PS1
PV3 refer to the header keywords \code{GAIN} and \code{RDNOISE}.  In
this case, the image is assumed to represent the readout from a single
detector, with well-defined gain and read noise characteristics.  This
assumption is not always valid.  For example, if the input flux image
is the result of an image stack with a variable number of input
measurements per pixel (due to masking and dithering), the variance
cannot be calculated from the signal image alone.  It is necessary in
such a case to supply a variance image which accurately represents the
variance as a function of position in the image.

Some image processing steps introduce cross-correlation between pixel
fluxes.  An obvious case is smoothing, but geometric transformations
which redistribute fractional flux between neighboring pixels also
introduces cross-correlations.  In the noise model, it is necessary to
track the impact of the cross correlations on the per-pixel variance.
In the general case, this would require a complete covariance image,
consisting of the set of cross-correlated pixels for each image pixel.
Since a typical smoothing or warping operation may introduce
correlation between 25 - 100 neighboring pixels, the size of such a
covariance image is prohibitive.  


Before sources are detected in the image, a model of the background is
subtracted.  The image is divided into a grid of background points
with a spacing defined by the \ippprog{psphot} recipe values
\code{BACKGROUND.XBIN, BACKGROUND.YBIN}, set to 400 pixels for PS1
PV3.  Superpixels of size \code{BACKGROUND.XSAMPLE, BACKGROUND.YSAMPLE}
($2 \times 2$ for PS1 PV3) times larger than
this spacing are used to measure the local background for each
background grid point, thus over-sampling the background spatial
variations.  In the interest of speed, a subset of \code{IMSTATS_NPIX}
(10,000 for PS1 PV3) randomly selected {\em unmasked} pixels in these
regions are used to determine the background.  The background value
for each superpixel is determined by fitting a Gaussian distribution
to the histogram of pixels values.  

If the image were empty of stars and only contained flux from a
uniform background sky, we would expect the distribution to be Poisson
distributed, and in general in a high-enough signal range to be
essentially Gaussian.  We fit a symmetric Gaussian to all histogram
bins within 15\% of the peak bin value to determine the mean and
standard deviation values for the background.  

If, however, the sky is not empty of stars or other sources, and we
have correctly masked the large majority of non-responsive pixels,
then we expect the flux distribution of the pixels to be asymmetric
with a Gaussian core representing the sky and a tail to the high end
representing the pixels with astronomical source flux contributions.
We would like to determine the mean of the underlying Gaussian without
suffering bias from the stellar flux.  We thus perform a second
Gaussian fit using an asymmetric subset of the histogram pixels,
fitting those histogram bins which are left of the peak but for which
the bin value is greater than 25\% of the peak bin, or right of the
peak but only using those bins for whch the bin value is greater than
50\% of the peak bin value.

If the fit to the asymmetric lower fraction of the curve is less than
the symmetric fit, but greater than the above lower-bound of the full
symmetric fit, then the lower fraction value is kept as the true mean
sky value for this superpixel.

Bilinear interpolation is used to generate a full-resolution image
from the grid of background points, and this image is then subtracted
from the science image.  The background image and the background
standard deviation image are kept in memory from which the values of
\code{SKY} and \code{SKY_SIGMA} are calculated for each source in the
output catalog.  For more details of the background subtraction, see
the discussion in Section~2.7 of \cite{waters2017}.


\subsection{Initial Source Detection}

\subsubsection{Peak Detection}
\label{sec:peaks}


The sources are initially detected by finding the location of local
peaks in the image.  The flux and variance images are smoothed with a
small circularly symmetric kernel using a two-pass 1D Gaussian.  The
smoothed flux and variance images are combined to generate a
significance image in signal-to-noise units, including correction for
the covariance, if known. At this stage, the goal is only to detect
the brighter sources, above a user defined S/N limit (configuration
keyword: \code{PEAKS_NSIGMA_LIMIT} = 20.0 for PS1 PV3).  A maximum of
\code{PEAKS_NMAX} (5000 of PS1 PV3) are found at this stage.  The
detection efficiency for the brighter sources is not strongly
dependent on the form of this smoothing function.

The local peaks in the smoothed image are found by first detecting
local peaks in each row.  For each peak, the neighboring pixels are
then examined and the peak is accepted or rejected depending on a set
of simple rules.  First, any peak which is greater than all 8
neighboring pixels is kept.  Any peak which is lower than any of the 8
neighboring pixels is rejected.  Any peak which has the same value as
any of the other 8 pixels is kept if the pixel $X$ and $Y$ coordinates
are greater than or equal to the other equal value pixels.  This
simple rule set means that a flat-topped region will result peaks at
the maximum $X$ and $Y$ corners of the region.

We use the 9 pixels which include the source peak to fit for the
position and position errors.  We model the peak of the sources as a
2D quadratic polynomial, and use a very simple bi-quadratic fit to
these pixels.  We use the following function to describe the peak

\[ f(x,y) = C_{00} + C_{10}x + C_{01} y + C_{11} x y + C_{20} x^2 + C_{02} y^2 \]

and write the Chi-Square equation:

\[ \chi^2 = \sum_{i,j} (F_{i,j} - f(x,y))^2 / \sigma_{i,j}^2 \]

By approximating the error per pixel as the error on just the peak,
and pulling that term out of the above equation, and recognizing that
the values x,y in the 3x3 grid centered on the peak pixel have values
of only 0 or 1, we can greatly simplify the chi-square equation to a
square matrix equation with the following values:
\[
\left( \begin{array}{cccccc}
9 & 0 & 0 & 0 & 6 & 6 \\ 
0 & 6 & 0 & 0 & 0 & 0 \\ 
0 & 0 & 6 & 0 & 0 & 0 \\ 
0 & 0 & 0 & 6 & 0 & 0 \\ 
6 & 0 & 0 & 0 & 6 & 4 \\ 
6 & 0 & 0 & 0 & 4 & 6 \\ 
\end{array} \right)
\left( \begin{array}{c}
C_{00}\\
C_{10}\\
C_{01}\\
C_{11}\\
C_{20}\\
C_{02}\\
\end{array} \right)
=
\left( \begin{array}{c}
\sum F_{i,j}     \\
\sum F_{i,j} x   \\
\sum F_{i,j} y   \\
\sum F_{i,j} x y \\
\sum F_{i,j} x^2 \\
\sum F_{i,j} y^2 \\
\end{array} \right)
\]

Inverting the 3x3 matrix terms for $C_{00}$, $C_{20}$, and $C_{02}$,
the location of the peak is determined from the minimum of the
bi-quadratic function above, and is given by:
\begin{eqnarray}
x_{min} & = & (C_{11} C_{01} - 2 C_{02} C_{10}) D^{-1} \\
y_{min} & = & (C_{11} C_{10} - 2 C_{20} C_{01}) D^{-1} \\
D      & = & 4 C_{20} C_{02} - C_{11}^2
\end{eqnarray}

The resulting peak position, ($x_{min}, y_{min}$), is used as the
default starting coordinate for the source.  Later in the
\ippprog{psphot} analysis, improved measurements of the source positions
are calculated as discussed below.

\begin{figure}[htbp]
  \begin{center}
 \includegraphics[width=\hsize,clip]{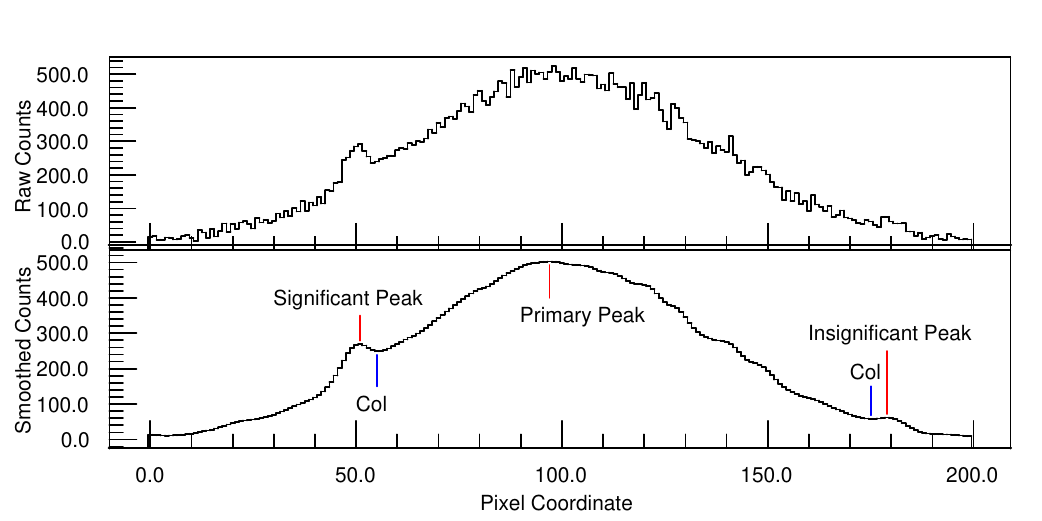}
  \caption{\label{fig:peaks} Illustration of peak finding and culling peaks within a
    footprint.  Insignificant peaks within the footprint of a brighter
    peak are ignored in further processing. }
  \end{center}
\end{figure}

\subsubsection{Footprints}
\label{sec:footprints}

The peaks detected in the image may correspond to real sources, but
they may also correspond to noise fluctuations, especially in the
wings of bright stars.  \ippprog{psphot} attempts to identify peaks which may be
formally significant, but are not locally significant.  It first
generates a set of ``footprints'', contiguous collections of pixels in
the smoothed significance image above the detection threshold
(\code{PEAKS_NSIGMA_LIMIT}).  These regions are grown by a small
amount to avoid errors on rough edges -- an image of the footprints is
convolved with a disk of radius \code{FOOTPRINT_GROW_RADIUS} (= 3
pixels for PS1 PV3).  Peaks are assigned to the footprints in which
they are contained (note by construction all peaks must be located in
a footprint since the peaks must be above the detection threshold).

For any peak which is not the brightest peak in that footprint it is
possible to reach the brightest peak by following the highest valued
pixels between the two peaks.  The lowest pixel along this path is the
{\em key col} for this peak (as used in topographic descriptions of a
mountain).  If the key col for a given peak is less than
\code{FOOTPRINT_CULL_NSIGMA_DELTA} (4.0 for PS1 PV3) sigmas below the
peak of interest, the peak is considered to be {\em locally
  insignificant} and removed from the list of possible detections (see
Figure~\ref{fig:peaks}).  In the vicinity of a saturated star, the
rule is somewhat more aggressive as the flat-topped or structured
saturated top of a bright star may appear as multiple peaks with
highly significant cols between them.  However, this is an artifact of
the proximity to saturation.  Sources for which the peak is greater
than 50\% of the saturation value require the col to also be a fixed
fraction (5\%) of the saturation below the peak to avoid being marked
as locally insignificant.

Sometimes it is useful to know if a source has a near neighbor which
may be affecting the photometry.  Three flag bits are used to identify
such possible situations.  Peaks which are {\em not} the brightest peak
within a single footprint have the flag bit
\code{PM_SOURCE_MODE2_HAS_BRIGHTER_NEIGHBOR} set.  This is a fairly
common situation.  We also define the following ratio to compare the
flux of the bright source to the flux of a neighbor scaled by
intervening area: $R = \frac{f_{\rm n}}{r^2 f_{\rm p}}$ where $f_{\rm
  n}$ is the flux of the brightest neighbor in the footprint, $f_{\rm
  p}$ is the flux of the source of interest, and $r$ is the separation
between the two sources.  If $R > 1$, the flag bit
\code{PM_SOURCE_MODE2_HAS_BRIGHT_NEIGHBOR_1} is set. If $R > 10$, the flag bit
\code{PM_SOURCE_MODE2_HAS_BRIGHT_NEIGHBOR_10} is set. 

\subsubsection{Centroid and Higher-Order Moments}
\label{sec:moments}

\begin{figure}[htbp]
  \begin{center}
  \includegraphics[width=0.95\hsize]{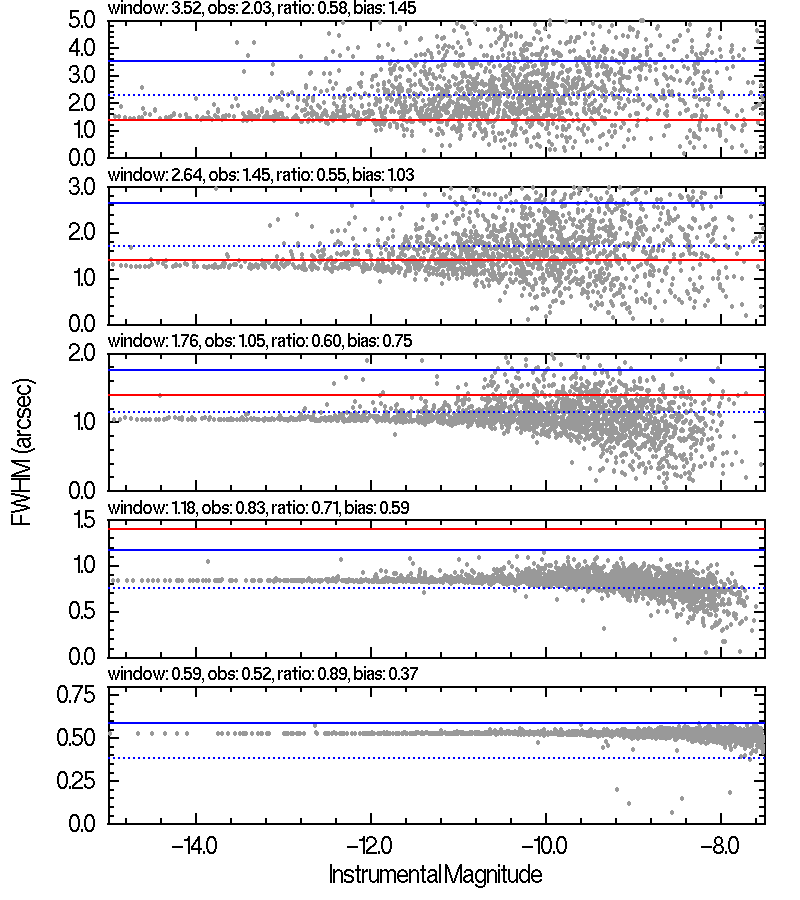}
  \caption{\label{fig:moments.window} Example of the biases
    encountered when measuring the second moments.  A simulated image
    was generated using the PS1 PSF profile.  Each panel corresponds
    to a different value of $\sigma_w$, corresponding to the window
    FWHM values as marked.  The solid red line is the true FWHM of the
    PSF used to generate the stars (1.4 arcsec in all cases).  The
    blue solid line is the FWHM of the window function.  The gray dots
    are the FWHM derived from the measured second moments for stars in
    the image.  The median of this distribution (mag $< -10$) is
    listed as ``obs''.  The ratio of the median FWHM to the FWHM of
    the window function is listed as ``ratio'', while the ratio of the
    median FWHM to the true stellar FWHM is listed as ``bias''.  The
    dotted blue line is the target (65\% of the window function).  In
    this example, we would choose $\sigma_w$ between 0.5 and 0.8
    arcseconds so the dotted blue line would match the bright end of
    the gray dots.   See discussion in the text for the choice of
    target window.
}
  \end{center}
\end{figure}

Once a collection of peaks has been identified, a number of basic
properties of the sources related to the first, second, and higher
moments are measured.  Below, the second moments are used to select
candidate stellar sources to be used in modeling the PSF.

In order to measure the moments, it is necessary to define an
appropriate aperture in which the moments are measured.  We also apply
a ``window function'', down-weighting the pixels by a Gaussian,
centered on the object, with size $\sigma_w$ chosen to be large
compared to the PSF size, $\sigma_{\rm PSF}$.  This window function
reduces the noise of the measurement of the moments by suppressing the
noisy pixels at high radial distance as well as by reducing the
contaminating effects of neighboring stars.  The choice of $\sigma_w$
and the aperture is an iterative process: for a given value of
$\sigma_w$, the PSF stars will have a measured value of the PSF size,
$\sigma^{\prime}_{\rm PSF}$ which different from the true value due to
the effect of the window function.  The measured value of the PSF size
will be biased high or low depending on both the signal-to-noise of
the source and the size of the window function compared to the true
PSF size.

These effects are illustrated in Figure~\ref{fig:moments.window} using
simulated data.  An image was generated with a PSF model matching the
radial profile of the PS1 PSF model with $\sigma_{\rm PSF}$
corresponding to a FWHM of 1.4 arcseconds.  As the window function
$\sigma_w$ is increased, the measured FWHM for the bright simulated
stars rises to meet the truth value.  For small values of $\sigma_w$,
fainter stars are biased to low measured values of the FWHM.  For
large values of $\sigma_w$, the faint stars are biased to higher
values and the scatter increases.  We attempt to minimize the scatter
and trends with magnitude at the cost of overall bias.

In a real image, we do not know the true value of the PSF size.  If we
simply choose a very large window function and rely on bright stars,
our estimate of the PSF size will be quite noisy.  Compounding this
problem are the two additional facts that (1) we do not know which are
the real stars (as opposed to bright galaxies or possible image
artifacts) and (2) the brighter stars are themselves subject to
additional biases due to saturation and other non-linear effects
\citep[c.f., ``the Brighter-Fatter''
  effect,][]{2014JInst...9C3048A,2015JInst..10C5032G}.  To make a
robust choice for $\sigma_w$, we choose a value such that the measured
value of $\sigma^{\prime}_{\rm PSF}$ is 65\% of $\sigma_w$.  The
resulting second moment values are biased somewhat low (\approx 75\%
of the truth value for the PS1 PSF profile), but are relatively
unbiased as a function of brightness.

To choose the value of $\sigma_w$, we try a sequence of values
spanning a range guaranteed to contain any reasonable seeing values.
The values are specified in the \ippprog{psphot} recipe as
\code{PSF.SIGMA.VALUES} and have the following values for PS1 PV3: (1,
2, 3, 4.5, 6, 9, 12, 18) pixels $\approx$ (0.26, 0.51, 0.77, 1.15,
1.54, 2.3, 3.1, 4.6) arcseconds.  For each of these $\sigma_w$ values,
we then select candidate PSF stars based on the distribution of the
measured $\sigma^{\prime}_{\rm PSF}$ in the two principal directions:
$\sigma_{x,x}$ and $\sigma_{y,y}$ (see
Section~\ref{sec:psf.source.selection}, below).  For each test value
of $\sigma_w$, we determine the ratio $\rho_\sigma =
\frac{\sigma_{x} + \sigma{y}}{2 \sigma_w}$, i.e., the ratio of the
window size to the observed PSF size.  We interpolate to find a value
of $\sigma_w$ for which $\rho_\sigma$ is expected to be 0.65.  We use
an aperture with a radius of 4$\sigma_w$ to select the pixels for the
measurement of the moments.


Once $\sigma_w$ has been determined, moments are measured as defined
below.

\begin{eqnarray}
x_0      & = & \frac{1}{S} \sum_i w_i (f_i - s_i)x_i \\
y_0      & = & \frac{1}{S} \sum_i w_i (f_i - s_i)y_i \\
M_{xx}   & = & \frac{1}{S} \sum_i w_i (f_i - s_i)(x_i - x_0)^2 \\
M_{xy}   & = & \frac{1}{S} \sum_i w_i (f_i - s_i)(x_i - x_0)(y_i - y_0) \\
M_{yy}   & = & \frac{1}{S} \sum_i w_i (f_i - s_i)(y_i - y_0)^2 \\
M_{xxx}  & = & \frac{1}{S} \sum_i \frac{w_i}{r_i} (f_i - s_i)(x_i - x_0)^3 \\
M_{xxy}  & = & \frac{1}{S} \sum_i \frac{w_i}{r_i} (f_i - s_i)(x_i - x_0)^2(y_i - y_0) \\
M_{xyy}  & = & \frac{1}{S} \sum_i \frac{w_i}{r_i} (f_i - s_i)(x_i - x_0)(y_i - y_0)^2 \\
M_{yyy}  & = & \frac{1}{S} \sum_i \frac{w_i}{r_i} (f_i - s_i)(y_i - y_0)^3 \\
M_{xxxx} & = & \frac{1}{S} \sum_i \frac{w_i}{r^2_i} (f_i - s_i)(x_i - x_0)^4 \\
M_{xxxy} & = & \frac{1}{S} \sum_i \frac{w_i}{r^2_i} (f_i - s_i)(x_i - x_0)^3(y_i - y_0) \\
M_{xxyy} & = & \frac{1}{S} \sum_i \frac{w_i}{r^2_i} (f_i - s_i)(x_i - x_0)^2(y_i - y_0)^2 \\
M_{xyyy} & = & \frac{1}{S} \sum_i \frac{w_i}{r^2_i} (f_i - s_i)(y_i - y_0)(y_i - y_0)^3 \\
M_{yyyy} & = & \frac{1}{S} \sum_i \frac{w_i}{r^2_i} (f_i - s_i)(y_i - y_0)^4
\end{eqnarray}
where $f_i$ is the flux in a pixel; $s_i$ is the local sky value for
that pixel; $w_i$ is the value of the window function for the pixel;
$S = \sum_i (f_i - s_i) w_i$ is the window-weighted sum of the source
flux, used to re-normalize the moments; $r_i$ is the radius of a
pixel, $\sqrt{(x_i - x_0)^2 + (y_i - y_0)^2}$; The sums are performed
over all (unmasked) pixels in the aperture.  For the centroid calculation ($x_0,
y_0$), the peak coordinate (see~\ref{sec:peaks}) is used to define the
aperture and the window function; for higher order moments, the
centroid is used to center the window function.

For sources with peak flux above the saturation limit, the moments are
generally poorly measured if the aperture defined by $\sigma_w$ is
used.  For these sources, the quality of the measurment is compromised
by the saturation.  However, it is still useful to estimate the first
and second moments of the source in order to allow a crude measurement
of the brightness from the wings of the source.  In this case, a
larger aperture, 3 times the standard aperture, is used to make a
crude estimate.  For such sources, the flag bit
\code{PM_SOURCE_MODE_BIG_RADIUS} is set and the source is ignored in
all analyses below except for the analysis applied to very bright
stars (Section~\ref{sec:very.bright.star}).

If the measured centroid coordinates ($x_0, y_0$) differ from the peak
coordinates be a large amount (1.5$\sigma_w$), then the peak is
identified as being of poor quality and is skipped in further
analyses; the flag bit
\code{PM_SOURCE_MOMENTS_FAILURE} is set for such sources.  In such
a case, it is likely that the `peak' was identified in a region of
flat flux distribution or many saturated or edge pixels.  During the
analysis of the moments, the background (``sky'') model is also examined for the
location of each source.  The value of the background and the variance
of the background are recorded for each source.  In some cases, the
sky model or the variance is not well defined at the location of a
specific sources (e.g., due to an extrapolation failure).  In these
cases, the flag bits \code{PM_SOURCE_SKY_FAILURE} or
\code{PM_SOURCE_SKYVAR_FAILURE} are set as appropriate and the
measurement of the moments is skipped.  

In addition to the moments above, the 1st and half-radial moments,
$M_r$ and $M_h$ as defined below, are calculated:
\begin{eqnarray}
M_r & = & \frac{1}{S} \sum_i (f_i - s_i)r_i \\
M_h & = & \frac{1}{S} \sum_i (f_i - s_i)\sqrt{r_i}
\end{eqnarray}
Note that the window function is not applied in the calculation of
these moments. 

With the first radial moment, we can calculate a preliminary Kron
radius and magnitude.  The Kron radius \citep{1980ApJS...43..305K} is
defined the be 2.5$\times$ the first radial moment.  The Kron flux is
the sum of (sky-subtracted) pixel fluxes within the Kron radius.  We
also calculate the flux in two related annular apertures: the Kron
inner flux is the sum of pixel values for the annulus $R_1 < r < 2.5
R_1$, while the Kron outer flux is the sum of pixel values for $2.5
R_1 < r < 4 R_1$.  The first radial moment is limited at the low and
high ends by $R_{\rm min} < M_r < R_{\rm max}$ where $R_{\rm min}$ is
the first radial moment of the PSF stars, or $0.75\sigma_w$ if that
cannot be determined.  $R_{\rm max}$ is set to the size of the moments
aperture, $4\sigma_w$.  These Kron measurements are performed for all
sources with a valid set of moments.  At this stage, the measurement
of the Kron parameters are preliminary since the aperture has been
chosen as a fixed size relative to the size of the PSF.  At a later
stage, higher-quality Kron parameters appropriate to galaxies are
measured with more care paid to the exact aperture used
(Section~\ref{sec:kron.mags}).


\subsection{PSF Determination}
\label{sec:PSF.Model}

\subsubsection{PSF Model vs Source Model}
\label{sec:Source.Model}

The point-spread-function (PSF) of an image describes the shape of all
unresolved sources in the image.  In a typical wide-field image, the
shape of unresolved sources varies as a function of position in the
image.  The full PSF thus needs to include a model with parameters
which vary across the image.

The PSF used by \ippprog{psphot} consists of an analytical function
combined with a pixelized representation of the residual differences
between the analytical model and the true PSF.  Both the shape
parameters of the analytical model and the pixelized residual
differences are allowed to vary in two dimensions across the images.

Within \ippprog{psphot}, several analytical models may be used to
describe the smooth portion of the PSF, but all share a few common
characteristics.  As an example, a simple model consists of a 2-D
elliptical Gaussian:
\begin{eqnarray}
f(x,y) & = & I_o e^{-z} + S  \\
    z  & = & \frac{x^2}{2\sigma_x^2} + \frac{y^2}{2\sigma_y^2} + \sigma_{\rm xy} x y \\
    x  & = & x_{\rm ccd} - x_o \\
    y  & = & y_{\rm ccd} - y_o 
\end{eqnarray}
Here the model parameters consist of the centroid coordinates ($x_o,
y_o$), the elliptical shape parameters ($\sigma_x, \sigma_y,
\sigma_{\rm xy}$), the model normalization ($I_o$) and the local value
of the background ($S$).  

A specific source will have a particular set of values for the model
parameters, some of which depend on the PSF model and the position of
the source in the image, while the rest are unique to the individual
source.  For the case of the elliptical Gaussian model, the PSF
parameters would be the shape terms ($\sigma_x, \sigma_y, \sigma_{\rm
  xy}$) while the independent parameters would be the centroid,
normalization and local sky values ($x_o, y_o, I_o, S$).  Thus the
shape parameters are each a function of the source centroid
coordinates:
\begin{eqnarray}
\sigma_x    & = & f_1(x_{\rm ccd},y_{\rm ccd}) \\
\sigma_y    & = & f_2(x_{\rm ccd},y_{\rm ccd}) \\
\sigma_{xy} & = & f_3(x_{\rm ccd},y_{\rm ccd}).
\end{eqnarray}
\ippprog{psphot} represents the variation in the PSF parameters as a function of
position in the image in two possible ways, specified by the
configuration.  The first option is to use a 2-D polynomial which is
fitted to the measured parameter values across the image.  The second
option is to use a grid of values which are measured for sources
within a subregion of the image.  In the latter case, the value at a
specific coordinate in the image is determined by interpolation
between the nearest grid points.  The order of the polynomial or the
sampling size of the grid is dynamically determined depending on the
number of available of PSF stars.  In the case of the PV3 analysis,
the grid of values was used, with a maximum of $6\times 6$ samples per
GPC1 chip image.  For the earlier PV2 analysis, the maximum grid
sampling was $3\times 3$ per GPC1 chip image.  For the PV1 analysis,
the polynomial representation was used, with up to 3rd order terms.
The higher order representation was used for PV3 in order to follow
some of the observed PSF variations in the images


Several analytical functions which are likely candidates to describe
the smooth portion of the PSF are available in \ippprog{psphot}:
\begin{itemize}
\item Gaussian : $f = I_0 e^{-z}$
\item Pseudo-Gaussian : $f = I_0 (1 + z + \frac{1}{2} z^2 + \frac{1}{6} z^3)^{-1}$ \code{[PGAUSS]}
\item Variable Power-Law : $f = I_0 (1 + z + z^{\alpha})^{-1}$ \code{[RGAUSS]}
\item Steep Power-Law : $f = I_0 (1 + \kappa z + z^{2.25})^{-1}$ \code{[QGAUSS]}
\item PS1 Power-Law : $f = I_0 (1 + \kappa z + z^{1.67})^{-1}$ \code{[PS1_V1]}
\end{itemize}
The Pseudo-Gaussian is a Taylor expansion of the Gaussian and is used
by Dophot \citep{1993PASP..105.1342S}.  The latter profiles are
similar to the Moffat profile form
\citep{1969AA.....3..455M,1983AA...126..278B}, with small differences.
A user may choose to try more than one analytical function for a given
image.  As discussed below (Section~\ref{sec:psf.model.choice}),
\ippprog{psphot} can automatically choose the best model based on the
quality of the PSF fits.

\begin{figure}[htbp]
  \begin{center}
  \includegraphics[width=\hsize]{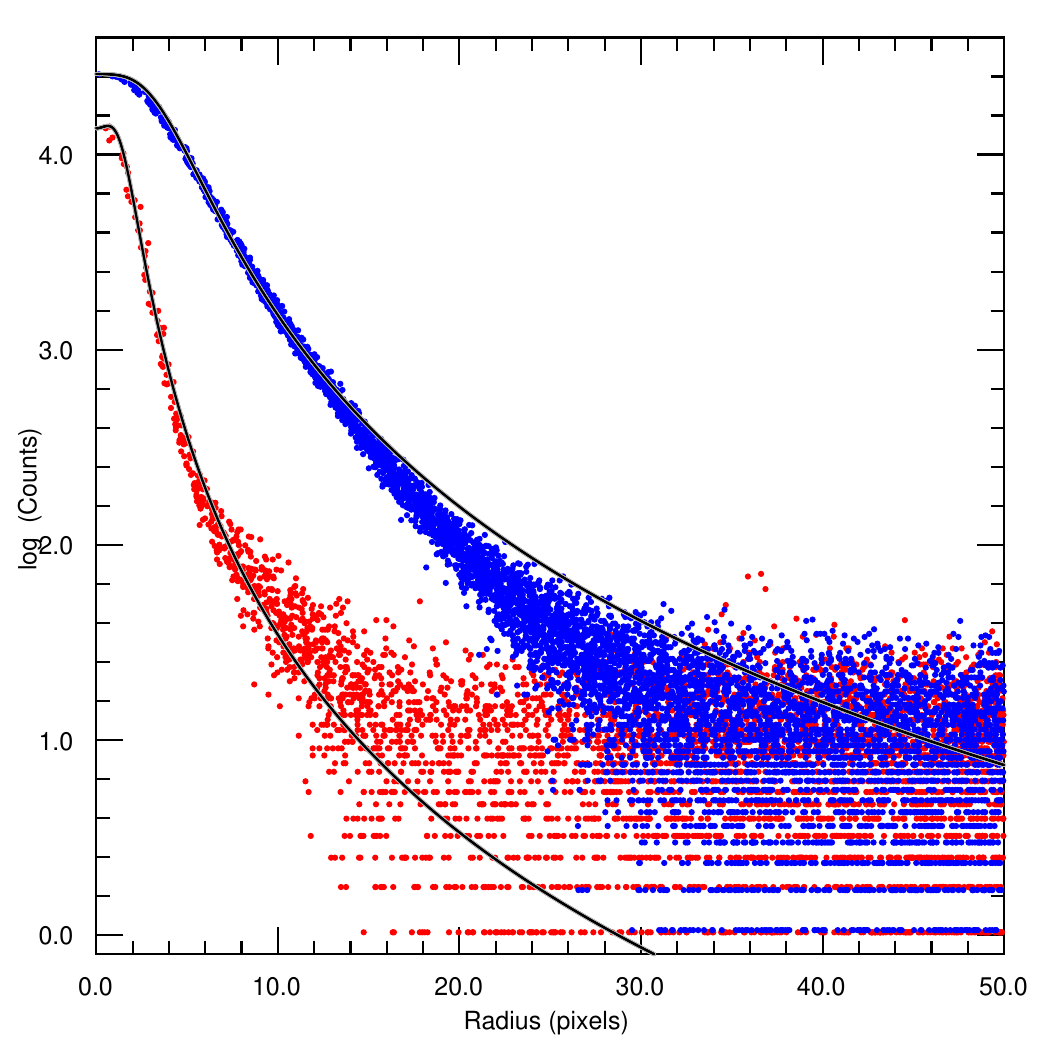}
  \caption{\label{fig:radial.profiles} Radial profiles of stellar images from PS1.  These two
    profiles illustrate the radial trend of the PS1 PSFs for a star
    with FWHM 0.9 arcsec (red) and 2.2 arcsec (blue).  The black line
    shows the PSF model with radial trend of the form $(1 + \kappa r^2 + r^{3.33})^{-1}$.}
  \end{center}
\end{figure}

For the PS1 GPC1 analysis, we used the \code{PS1_V1} model, which we
found by experimentation to match well to the observed profiles
generated by PS1.  Figure~\ref{fig:radial.profiles} shows example
radial profiles for moderately bright stars in fairly good (0.9
arcsec) and poor (2.2 arcsec) seeing.  Using a fixed power-law
exponent results in somewhat faster profile fitting compared to the
variable power-law exponent model.

The analytical models in \ippprog{psphot} are written with a high degree
of code abstraction making it relatively easy to add different
analytical models to the software.  The same portion of code used to
describe the analytical portion of the PSF sources is also used to for
galaxy models. 


Once the smooth component of the PSF has been fitted with an
analytical model, a pixel representation of the residuals is
generated.  This representation is constructed as an image of the
expected residuals for any position in the image.  The value of each
pixel in the image model is determined from 2D fits to the measured
residuals of the PSF stars.  

The residual model is calculated using the residuals for all PSF
stars.  The residuals (and their errors) for each star are
renormalized by the flux of the star to put them on a consistent flux
scale.  For each PSF star, all pixels within a user-specified radius
(\code{PSF.RESIDUALS.RADIUS = 9}) are selected for the measurement.  For a
given pixel in the model, the pixel values from the PSF stars are
interpolated to the center of the model pixel. Pixels may be used in
this analysis if their signal-to-noise exceeds a user-defined limit.
For the PV3 $3\pi$ analysis, we allowed all pixels within the
user-specified radius, not limiting on the basis of the
signal-to-noise.

Pixels for a given star which are more than a number of sigmas
(\code{PSF.RESIDUALS.NSIGMA = 3.0}) deviant from the median value of
the pixels from all stars are rejected.

If no spatial variation is allowed, the mean or median value is
calculated for the model pixel based on the user-specified mean
statistic (\code{PSF.RESIDUALS.STATISTIC = ROBUST_MEDIAN}).

If spatial variation is requested, then the pixel values are fitted to
a linear model:
\[
\begin{array}{lll}
R[(x_{\rm mod},y_{\rm mod})][(x_{\rm ccd},y_{\rm ccd})] & = & R_o[(x_{\rm mod},y_{\rm mod})] \\
& + & R_x[(x_{\rm mod},y_{\rm mod})] x_{\rm ccd} \\
& + & R_y[(x_{\rm mod},y_{\rm mod})] y_{\rm ccd} \\
\end{array}
\]
where $R[(x_{\rm mod},y_{\rm mod})][(x_{\rm ccd},y_{\rm ccd})]$ is the
value for model pixel $(x_{\rm mod},y_{\rm mod})$ for a star with
centroid at image pixel $(x_{\rm ccd},y_{\rm ccd})$.  The parameters
$R_o, R_x, R_y$ are determined for each pixel in the model $[(x_{\rm
    mod},y_{\rm mod})]$.

\subsubsection{Candidate PSF Source Selection}
\label{sec:psf.source.selection}

\begin{figure}[htbp]
  \begin{center}
  \includegraphics[width=\hsize]{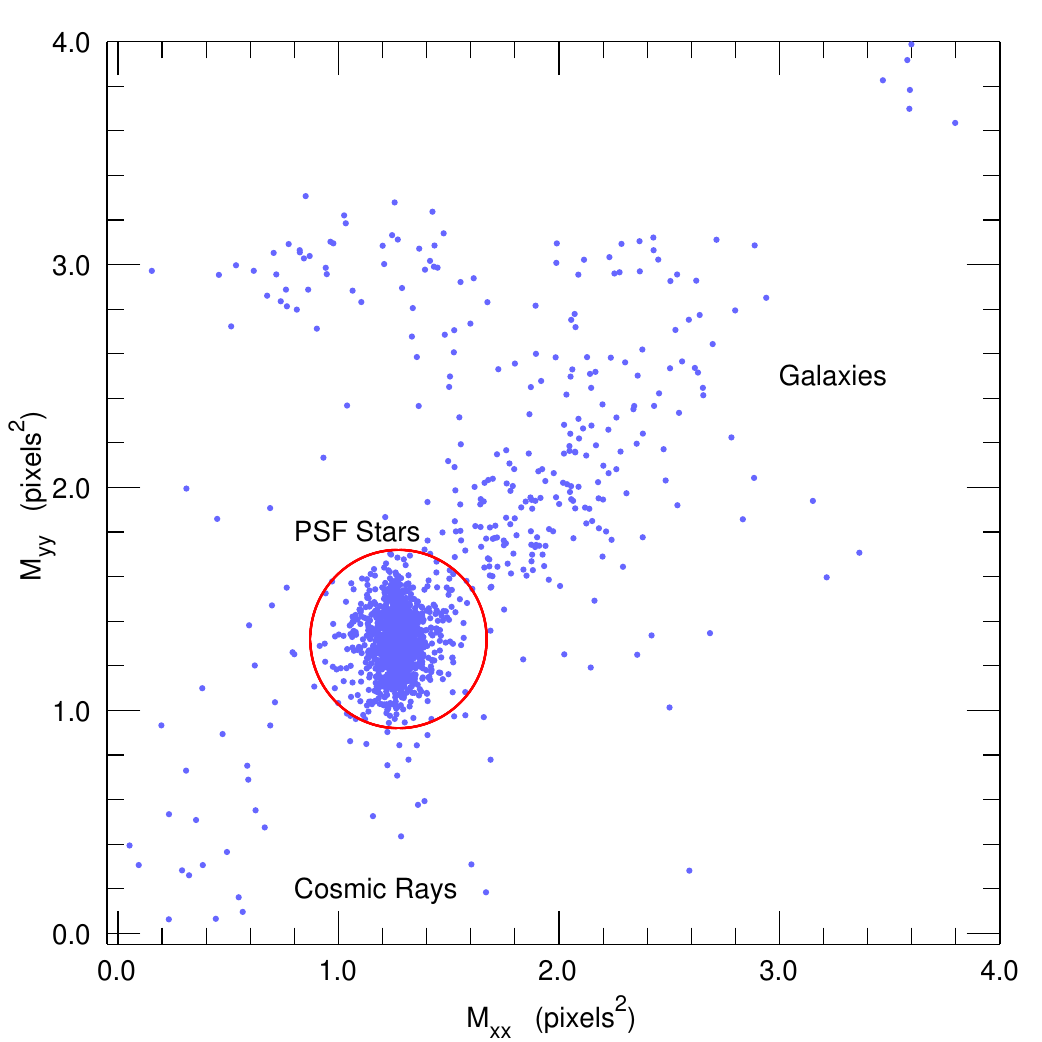}
  \caption{\label{fig:moment.class} Illustration of PSF star selection
    using the second moments in $X_{\rm ccd}$ and $Y_{\rm ccd}$
    directions.  The dominant clump is located in this diagram.
    Galaxies tend to have a range of sizes and thus spread out above
    the stars.  Cosmic rays also have a range of sizes, with one
    dimension smaller than the PSF.  The red circle represents the PSF
    star candidates. }
  \end{center}
\end{figure}

The first stage of determining the PSF model for an image is to
identify a collection of sources in the image which are {\em likely}
to be unresolved (i.e., stars).  \ippprog{psphot} uses the source sizes as
estimated from the second moments to make the initial guess at a
collection of unresolved sources.  At this point, the program has
measured the second order moments for all sources identified by their
peaks, as well as an approximate signal-to-noise ratio, above the
bright threshold.  All sources with a S/N ratio greater than a
user-defined parameter (\code{PSF_SN_LIM} = 20.0 for PS1 PV3) are
selected by \ippprog{psphot}, though sources which have more than a
certain number of saturated pixels are excluded at this stage.  The
program then examines the 2-D plane of $M_{x,x}, M_{y,y}$ in search
of a concentrated clump of sources (see
Figure~\ref{fig:moment.class}).  To do this, it constructs an
artificial image with pixels representing the value of $M_{x,x},
M_{y,y}$, using $0.1 \sigma^2_w$ as the size of a pixel in this
artificial image.  The binned $M_{x,x}, M_{y,y}$ plane is then
examined to find a significant peak.  Unless the image is extremely
sparse, such a peak will be well-defined and should represent the
sources which are all very similar in shape.  Other sources in the
image will tend to land in very different locations, failing to
produce a single peak.  To avoid detecting a peak from the unresolved
cosmic rays, sources which have second-moments very close to 0 are
ignored.  For these sources, the flag bit \code{PM_SOURCE_MODE_DEFECT}
is set.

Once a peak has been detected in this plane, the centroid and second
moments of this peak are measured.  All sources which land within 2
pixels of this centroid are selected as candidate PSF sources in the
image.

When the second moments are measured, \ippprog{psphot} also counts the
number of saturated pixels within the analysis aperture.  If more than
a single saturated pixel is found, and if the second moments of that
object are more than one standard deviation larger than the clump
identified above, this source is identified as a highly saturated star
and marked with the flag bit \code{PM_SOURCE_MODE_SATSTAR}.  Sources
which have more than a single saturated pixel, but for which the
second moments do not exceed the above limits are marked as likely
saturated regions (e.g., bleed trails).  These sources are skipping in
most additional analyses and are marked with the flag bit
\code{PM_SOURCE_MODE_SATURATED}.

\subsubsection{Candidate PSF Source Model Fits}
\label{sec:psf.model.choice}


All candidate PSF sources are then fitted with the selected source
model, allowing all of the parameters (PSF and independent) to vary in
the fit.  The software uses the Levenberg-Marquardt minimization
technique \citep{1992nrca.book.....P,Madsen} for the non-linear fitting.  Non-linear
fitting can be very computationally intensive, particularly if the
starting parameters are far from the minimization values.  The first
and second moments are used to make a good guess for the centroid and
shape parameters for the PSF models.  Any sources which fail to
converge in the fit are flagged as invalid.

For the resulting collection of source model parameters, the
PSF-dependent parameters of the models are all fitted as a function of
position using either the 2-D polynomial or the gridded superpixel
representation.  The maximum order of these fits depends on the number
of PSF sources (see Table~\ref{tab:psf.order.nstars}).  The fitting process for
these polynomials is iterative, and rejects the $3\sigma$ outliers in
each of three passes.  This fitting technique results in a robust
measurement of the variation of the PSF model parameters as a function
of position without being excessively biased by individual sources
which are not well described by the PSF model (e.g., galaxies which
snuck into the sample).  Sources whose model parameters are rejected
by this iterative fitting technique are also marked as invalid PSF
sources and ignored in the later PSF model fitting stages.  Sources
which are actually used to define the PSF model for a given image have
the flag bit \code{PM_SOURCE_MODE_PSFSTAR} set.

The order of the fit or number of grid samples is modified if the
number of stars available for the fit is insufficient to justify the
highest value.  Regardless of the requested order, if the number of
stars is below the following limits, the order is limited as shown in
Table~\ref{tab:psf.order.nstars}.  Note that the number of grid cells
in one dimension is one greater than the equivalent polynomial order.

\begin{table}
\caption{\label{tab:psf.order.nstars} Minimum number of stars required
  for a given order of the PSF 2D variations.} 
\begin{center}
\begin{tabular}{lll}
\hline
\hline
{\bf Minimum Number} & {\bf Order} & {\bf Number of} \\
{\bf of Stars}       &             & {\bf Grid Cells} \\
\hline
 16 &  1 &  4 \\
 54 &  2 &  9 \\
128 &  3 & 16 \\
300 &  4 & 25 \\
576 &  5 & 36 \\
\hline
\end{tabular}
\end{center}
\end{table}

All of the PSF-candidate sources are then re-fitted using the PSF
model to specify the PSF-dependent model parameter values for each
source.  For example, in the case of the elliptical Gaussian model,
the shape parameters ($\sigma_x, \sigma_y, \sigma_{xy}$) for each
source are set by the coordinates of the source centroid and fixed
(not allowed to vary) in the fitting procedure.  The resulting fitted
models are then used to determine a metric which tests the quality of
the PSF model for this particular image.

The metric used by \ippprog{psphot} to assess the PSF model is the
scatter in the differences between the aperture and fit magnitudes for
the PSF sources.  This difference is a critical parameter for any PSF
modeling software as it is a measurement of how well the PSF model
captures the flux of the star.  Aperture photometry is measured for a
circular aperture with a radius of \code{PSF_APERTURE_SCALE} (= 4.5
for the PV3 $3\pi$ analysis) times $\sigma_w$
(Section~\ref{sec:moments}).  The average aperture correction ($m_{\rm
  AP} - m_{\rm PSF}$) is measured and, if multiple PSF model types are
selected, the PSF model with the minimum clipped scatter in this
statistic is chosen for the image.  An approximate aperture correction
is measured here, with a more detailed correction measured after all
source analysis is performed (see
Section~\ref{sec:aperture.correction}).  Sources for which the
aperture magnitude is measured have the flag bit
\code{PM_SOURCE_MODE_AP_MAGS} set.  These aperture magnitudes are
stored in the DVO field \code{Measure.Map} and exported to the PSPS as
a flux in Janskies in the field \code{Detection.apFlux}.  The radius
(in arcseconds)
of the aperture used for each exposure is reported in PSPS as
\code{Detection.apRadius}, while the unmasked fraction of the aperture
is reported in PSPS as \code{Detection.apFillF}.

When the PSF and aperture photometry for a source is measured, two
additional quantities are measured which are useful to assess the
quality of the measurements.  First, the mask image is examined and the
number of unmasked pixels is summed, weighted by the normalized PSF
model.  The resulting quantity, \code{PSF_QF} has a value between 0.0
(totally masked) and 1.0 (totally unmasked).  Elsewhere in the IPP
system, we use this value to filter out detections which are
unreliable due to the masking.  For a generous cut, leaning toward
completeness at the cost of some lower quality measurements,
\code{PSF_QF} $> 0.85$ is used in some contexts; in other cases, we
require \code{PSF_QF} $> 0.95$ to ensure a high-quality measurement
\citep[see for example the calculation of average photometry
  in][]{magnier2017.calibration}.  The second quantity is related to
the first: \code{PSF_QF_PERFECT} uses all mask values to assess the
quality factor, while \code{PSF_QF} uses only the ``bad'' mask bit
values (see Section~\ref{sec:image.preparation}).

Several flag bits are raised based on statistics which are similar to
the \code{PSF_QF} measurement.  First, \ippprog{psphot} calculates the
normalized, PSF-weighted fraction of pixels which are masked due to
one of the following four causes: a diffraction spike (\code{SPIKE}),
the core of a saturated star (\code{CORE}), burntool-subtracted region
(\code{BURNTOOL}), or a pixel for which, due to interpolation or
convolution, a significant fraction of the pixel flux comes from a
masked pixel.  These masking conditions are all treated as ``suspect''
by \ippprog{psphot}, which means they are {\em included} in the analysis of
the source pixels.  However, since they may potentially affect the
photometry (or astrometry), it is useful to note of a source has a
non-trivial fraction of these poor mask pixels.  If the normalized
PSF-weighted fraction of pixels masked due to any of these four
conditions exceeds 25\%, then one of the following bits is raised for
the corresponding condition:  \code{PM_SOURCE_MODE2_ON_SPIKE},
\code{PM_SOURCE_MODE2_ON_STARCORE},
\code{PM_SOURCE_MODE2_ON_BURNTOOL},
\code{PM_SOURCE_MODE2_ON_CONVPOOR}.  In addition, the following flag
bits may also be raised if the central pixel
of a source lands on a pixel masked for a diffraction spike
(\code{PM_SOURCE_MODE_ON_SPIKE}), an optical ghost
(\code{PM_SOURCE_MODE_ON_GHOST}), or off the active pixels of the CCD (\code{PM_SOURCE_MODE_OFF_CHIP}).

\subsection{Bright Source Analysis}

Once a PSF model has been determined, the brighter sources in the
image may be analyzed in detail.  The goals in this stage are (1) to
determine the fluxes and positions of the bright stellar sources with
high precision appropriate to their high signal-to-noise and (2) to
characterize the bright source flux profiles sufficiently well that
they may be subtracted from the image to allow for the clean detection
of the fainter sources.  Note that as the analysis proceeds, there are
several stages in which the 2D flux models for all sources are
subtracted from the image, and individual sources are replaced in the
image for a particular analysis step and then removed again.  The flux
limit for this analysis stage is user-defined as a signal-to-noise
value.  In the PV3 analysis of the $3\pi$ survey data, this limit was
set to a signal-to-noise ratio of 20.0.

In order to allow for multiple threads to process a single image, the
pixels in an image are divided into a grid of superpixels.  The
superpixels are assigned to one of four groups so that each superpixel
in a group is well separated from the other superpixels of that group.
The analysis of the image proceeds in 4 steps, one for each of these
groups.  Each of the superpixels in the first group is assigned to a
single thread until all threads are assigned.  A single thread is
responsible for the analysis of sources which land within their
current superpixel, as determined by the centroid coordinates.  Since
the superpixels in a given thread group are not contiguous by
construction, sources near the edge of a superpixel can be analysed by
considering the nearby pixels from neighboring superpixel (guaranteed
not to be in the current thread group).

As the threads complete their analysis, they are assigned the next
unfinished superpixel in the active group.  When all superpixels in
one group have been processed, then the superpixels in the next group
can start.  This strategy allows the threading to process sources
which may be extended without the danger that two threads are actively
touching the same pixels.  For the PV3 analysis, 4 threads were used
for most processing tasks.

\subsubsection{Very Bright Stars}
\label{sec:very.bright.star}

The standard \ippprog{psphot} PSF modeling code fails to fit the wings of
highly saturated stars, especially if the core of the star is too
contaminated by saturated pixels.  For stars with more than a single
saturated pixel, we model the radial profile of the logarithmic
instrumental flux in logarithmically spaced radial bins.  For each
radial bin, we determine the median of the log-flux.  This median
profile is then interpolated to generate the full radial flux
distribution.  Note that in the case of very saturated stars, pixels
in the central regions are largely masked, because they are
saturated.  Thus in these cases, the psf-weighted masked fraction (see
Section~\ref{sec:psf.model.choice}) is generally quite low or 0.0.
Sources for which this radial profile is subtracted have the flag bit
\code{PM_SOURCE_MODE2_SATSTAR_PROFILE} set.


\subsubsection{Fast Ensemble PSF Fitting}

Before the detailed analysis of the sources is performed, it is
convenient to subtract off all of the sources, at least as well as
possible at this stage.  We make the assumption that all sources are
PSF-like.  If the centroid of the source has been determined, we use
this value for its position; otherwise, we use the interpolated
position of the peak. A single linear fit is used to simultaneously
measure all source fluxes.  Since the local sky has been subtracted,
this measurement assumes the local sky is zero.  We can write a single
$\chi^2$ equation for this image:
\[
\chi^2 = \sum_{\rm pixels} (F_{x,y} - \sum_{\rm sources} A_i P[x_0,y_0])^2
\]
where $F_{x,y}$ is image flux for each pixel, $P[x_0,y_0]$ is the PSF
model realized at the position of source $i$, and $A_i$ is the
normalization for the source.

Minimizing this equation with respect to each of the $A_i$ values
results in a matrix equation:
\[ M_{i,j} \bar{A_i} = \bar{F_j}\]
where $\bar{A_i}$ is the vector of $A_i$ values, the elements of
$M_{i,j}$ consist of the dot products of the unit-flux PSF for source
$i$ and source $j$, and $\bar{F_j}$ is the dot product of the
unit-flux PSF for source $j$ with the pixels corresponding to source
$j$.  The dot products are calculated only using pixels within the
source apertures.  Since most sources have no overlap with most other
sources, this matrix equation results in a very sparse, mostly
diagonal square matrix.  The dimension is the number of sources,
likely to be 1000s or 10,000s.  Direct inversion of the matrix would
be computationally very slow.  However, an iterative solution quickly
yields a result with sufficient accuracy.  In the iterative solution,
a guess at the solution $\bar{A}$ is made assuming $M_{i,j}$ is purely
diagonal; the guess is multiplied by $M_{i,j}$, and the result
compared with the observed vector $\bar{F_j}$.  The difference is used
to modify the initial guess.  This process is repeated several times
to achieve convergence.  Convergence is quick (a few iterations)
because of the highly diagonal matrix with small off-diagonal terms:
the dot product of source $i$ and source $j$ is 1 where $i = j$ and
much less than 1 where $i \neq j$.

Once a solution set for $A_i$ is found, all of the sources are
subtracted from the image by applying these values to the unit-flux
PSF.  Sources for which a PSF model has been fitted (whether or not
this is retained as the best model in the end) has the flag field
\code{PM_SOURCE_MODE_PSFMODEL} set.  All sources which are included in
the ensemble linear fit have the flag bit
\code{PM_SOURCE_MODE_LINEAR_FIT} set, including those for which the
model is not the PSF.

\subsubsection{Radial Profile Wings}
\label{sec:radial.profile}

We attempt to measure the radial profile of sources in order to find
the radius at which the flux of the source is matches the sky.  In
this analysis, a series of up to 25 radial bins with power-law spacing
are defined and the flux of the source in each annulus is measured.
The ``sky radius'' is defined to be the radius at which the (robust
median) flux in the annulus is within 1 $\sigma$ of the local sky
level.  If this limit is not reached before the slope of the flux from
one annulus to the next is less than a user-defined limit, then the
annulus at which the slope reaches this limit is used to define the
sky radius.  These values are saved in the output smf / cmf files, but
not sent to the PSPS.  The sky radius value is used below in the
calculation of the Kron magnitude.

\subsubsection{Kron Magnitudes}
\label{sec:kron.mags}

Preliminary Kron radius and flux values \citep{1980ApJS...43..305K}
are calculated soon after sources are detected
(Section~\ref{sec:moments}).  However, these preliminary values are
not accurate due to the window-functions applied.  After sources have
been characterized and the PSF model is well-determined, the Kron
parameters are re-calculated more carefully.  In this version of the
calculation, following the algorithm described by \cite{sextractor},
the image is first smoothed by Gaussian kernel with $\sigma = 1.7$
pixels, corresponding to a FWHM of 1.0\arcsec\ in the PS1 stack
images.  Next, the Kron radius is determined in an iterative process:
the first radial moment is measured using the pixels in an aperture
6$\times$ the first radial moment from the previous iteration.  On the
first iteration, the sky radius is used in place of the first radial
moment.  By default, 2 iterations are performed.  The Kron radius is
defined to be 2.5$\times$ the first radial moment.  The Kron flux is
the sum of pixel fluxes within the Kron radius.  We also calculate the
flux in two related annular apertures: the Kron inner flux is the sum
of pixel values for the annulus $R_1 < r < 2.5 R_1$, while the Kron
outer flux is the sum of pixel values for $2.5 R_1 < r < 4 R_1$.

Two details in the calculation above should be noted.  First, for
faint sources, noise in the measurement of the 1st radial moment may
result in an excessively small aperture for the successive
calculations.  The window used for the calculations is constrained to
be at least the size of the aperture based on the PSF stars
(Section~\ref{sec:moments}).  At the other extreme, noise may make the radius
grow excessively, resulting in an unrealistically low effective
surface brightness.  The aperture is constrained to be less than a
maximum value defined such that the minimum surface brightness is
1/2$times$ the effective surface brightness of a point source detected at the
$5\sigma$ limit.

Second, the measurement of the 1st radial moment includes a filter to
reduce contamination from outlier pixels.  Pairs of pixels on
opposites sides of the central pixel are considered together.  The
geometric mean of the two fluxes is used to replace the flux values.
If the source has 180\degree\ symmetry, this operation has no impact.
However, if one of the two pixels is unusually high, the value will be
suppressed by the matched pixel on the other side.  This trick has the
effect of reducing the impact of pixels which include flux from near
neighbors.


\subsubsection{Source Size Assessment}
\label{sec:source.size}

After the PSF model has been fitted to all sources, and the Kron flux
has been measured for all sources, \ippprog{psphot} uses these two
measurements, along with some additional pixel-level analysis, to
determine the size class of the source.  Sources identified as
extended will be fitted with a galaxy model (or possibly another type
of extended source model in special cases).  If the source is small
compared to a PSF, it is considered to be a {\em cosmic ray} and
masked.

Extended sources are identified as those for which the Kron magnitude
is significantly brighter than the PSF magnitude when compared to a
PSF star.  The value $\delta M_{rm KP} = m_{\rm Kron} - m_{\rm PSF}$,
the difference between the PSF and Kron magnitudes, is calculated for
each source.  The median of $\delta M_{rm KP}$ is calculated for the
PSF stars.  This median is subtracted from $\delta M_{rm KP}$ for each
star.  The result is divided by the quadrature error of the PSF and
Kron magnitudes and called \code{extNsigma}.  If \code{extNsigma} is
larger than \code{PSPHOT.EXT.NSIGMA.LIMIT} (3.0), the source is
considered to be extended and the flag bit
\code{PM_SOURCE_MODE_EXT_LIMIT} is set for the source.

Cosmic rays are identified by a combination of the Kron magnitude and
the second-moment width of the source in the minor axis direction.
The second-moment in the minor axis direction is calculated from
$M_{xx}, M_{xy}, M_{yy}$ as follows:
\[
M_{\rm minor} = \frac{1}{2}(M_{xx} + M_{yy}) - \frac{1}{2}\sqrt{(M_{xx} - M_{yy})^2 + 4 M_{xy}^2}
\]
If $M_{\rm minor} < 0.8$ pixels$^2$ and the signal-to-noise of the
flux measured in the moments analysis $> 7$, then the source is
identified as a cosmic ray and the associated pixels are masked.
These values are tuned empirically for the PV3 analysis based on
cosmic rays identified in the GPC1 images.  Sources which are
determined to the a cosmic ray in this manner have the flag bit
\code{PM_SOURCE_MODE_DEFECT} set.

The pixels of any suspected cosmic ray identified above are examined
in additional detail to make a final judgement.  The Laplacian edge
detection algorithm based on \cite{2001PASP..113.1420V} is used to
check for sharp edges in the flux distribution.  If the sharpness
exceeds a defined limit, then the pixels are masked and the flag bit
\code{PM_SOURCE_MODE_CR_LIMIT} is set for the source.



\subsubsection{Full PSF Model Fitting}
\label{sec:nonlinear.psf.model}



Once a PSF model has been selected for an image, \ippprog{psphot}
attempts to fit all of the detected sources, with signal-to-noise
ratio greater than a user-defined limit, with the PSF model.  In the
PV3 analysis of the $3\pi$ survey data, this limit was set to a
signal-to-noise ratio of 20.0 for all analysis stages.  In these fits,
the dependent parameters are fixed by the PSF model and only the 4
independent source model parameters are allowed to vary in the fit.
\ippprog{psphot} again uses Levenberg-Marquardt minimization for the
non-linear fitting.  The sources are fitted in their S/N order,
starting with the brightest and working down to the user-specified
limit, with the other sources subtracted as discussed above.  All
sources for which a non-linear PSF model has been attempted have the
flag bit \code{PM_SOURCE_MODE_FITTED} set, regardless of the quality
of that fit.

Since the PSF model describes the variation of the PSF across the
image, the parameters used to fit a specific object are drawn from the
model at the position corresponding to the object centroid.
Occasionally, a PSF model for an image may not be well determined in
all regions of the image.  For example, not enough bright stars were
available across the full range of the image to model the PSF and the
resulting fitted parameters yield non-sensical solutions in areas
where detected (fainter) sources are found.  In such cases, the PSF
fitting is skipped and the flag bit \code{PM_SOURCE_MODE_BADPSF} is set.

For the PSF model fitting, only pixels within a circular aperture
scaled based on the seeing are used.  The radius of the circular
aperture is set to be a fixed multiple (\code{PSF_FIT_RADIUS_SCALE})
of $\sigma_w$, the width of the Gaussian window function determined
based on the analysis of the second moments (see
Section~\ref{sec:moments}).  For the PV3 $3\pi$ analysis, the PSF fit
window radius is $7 \times \sigma_w$.

Sources which are blended with other sources may be fitted together as a
set of PSFs.  Blended objects are identified by first searching for
objects for which the PSF fit windows overlap.  For a group of such
neighboring objects, a contour is determined in the flux image at
$25\%$ of the peak of the brightest source in the group.  All objects
lying within this contour are treated as blends of this brightest
source.  If other objects in this group exist, the brightest object
not already assigned to a blend is used to define the contour for
blends of this next object.  All objects in the image are tested as
possible blends.  A single multi-source fit is performed on each group
of blended peaks.  Sources which are identified as members of a
blended group have the flag bit \code{PM_SOURCE_MODE_BLEND} set, while
those for which a blended PSF fit succeeds have the flag bit
\code{PM_SOURCE_MODE_BLEND_FIT} set.  {\em Note that for DR1 \& DR2,
  this option was not used because it tended to prevent galaxies from
  being fitted as extended objects; the rules for identifying blended
  stars and galaxies will be revisited in future re-analyses.}

After the PSF model is fitted to each object, \ippprog{psphot} makes an
assessment of the quality of the PSF fits.  First, it checks that the
non-linear fitting process has converged with a valid fit.  The fit
for an object can fail if there are too few valid pixels, due to
masking or proximity to an edge, or if the parameters are driven to
extreme values which are not permitted.  In addition, it is possible
for the peak finding algorithm to identify peaks in locations which
are not actually a normal peak.  Some of these cases are in the edges
of saturated, bleeding columns from bright stars, in the nearly flat
halos of very bright stars, and so on.  In these cases, a local peak
exists, with a lower nearby sky region.  However, the fitted PSF model
cannot converge on the peak because it is very poorly defined (perhaps
only existing in the smoothed image).  In these cases, \ippprog{psphot}
flags the object with the bad bit \code{PM_SOURCE_MODE_FAIL}.  It is
also possible in this type of case for the fit to result in a very low
or negative value for the flux normalization parameter.  Source for
which the peak is less than 0.02 are also marked as failing the
non-linear PSF fit (\code{PM_SOURCE_MODE_FAIL}).

Poor fits are also identified by the signal-to-noise and the $\chi^2$
value of the resulting fit.  If a source has a PSF S/N ratio lower
than a user-defined cutoff (set to 2.0 for the PV3 analysis of the
$3\pi$ survey), the non-linear PSF fit will be rejected.  If the
Chi-Square per degree of freedom is greater than a user-defined limit
(set to 50.0 for the PV3 analysis of the $3\pi$ survey), the
non-linear PSF fit will be rejected.  These sources are marked with
the flag bit (\code{PM_SOURCE_MODE_POOR}).

Sources which are pass the above tests are marked as having a valid
non-linear PSF model fit with the flag bit
\code{PM_SOURCE_MODE_NONLINEAR_FIT}.  Among these sources, those for
which the peak flux is greater than the saturation limit (see
Section~\ref{sec:image.preparation}) are marked as saturated stars
(\code{PM_SOURCE_MODE_SATSTAR}).  These model fits should be
considered with caution, but the fluxes and positions may have some
validity.


As the sources are fitted to the PSF model, those which survive the
exclusion stage are subtracted from the image.  The subtraction
process modifies the image pixels (removing the fitted flux, though
not the locally fitted background) but does not modify the mask or the
variance images.  The signal-to-noise ratio in the image after
subtraction represents the significance of the remaining flux.  If the
subtractions are sufficiently accurate models of the PSF flux
distribution, the remaining flux should be below 1 $\sigma$
significance.  In practice the cores of bright stars are poorly
represented and may have larger residual significance.

For sources in groups of blended stars, the resulting fits are
evaluated independently.  Any which are determined to be valid PSF
fits are subtracted from the image and kept for future analysis.

\subsubsection{Double and Extended Sources}

Sources which are judged to be non-PSF-like are confronted with two
possible alternative choices.  First, the source is fitted with a
double-source model.  In this pass, the assumption is made that there
are two neighboring sources, but the peaks are not resolved.  The
initial guess for the two peaks is made by splitting the flux of the
single source in half and locating the two starting peaks at +/- 2
pixels from the original peak along the direction of the semi-major
axis of the sources, as measured from the second moments.  In order
for the two-source model to be accepted, both sources must be judged
as a valid PSF source.  Otherwise, the double-PSF model is rejected
and the source is fitted with the available non-PSF model or models.
Sources for which a double-PSF model is fitted have the flag bit
\code{PM_SOURCE_MODE_PAIR} set.

\subsubsection{Non-PSF Sources}
\label{sec:nonlinear.galaxy.model}

Once every source (above the S/N cutoff) has been confronted with the
PSF model, the sources which are thought to be extended (resolved) can
now be fit with an appropriate model (e.g., galaxy profile or other
likely extended shapes).  Again, the fitting stage starts with the
brightest sources (as judged by the rough S/N measured from the
moments aperture) and working to a user defined S/N limit.

\ippprog{psphot} will use the user-selected extended source model to
attempt these fits.  In the configuration system, the keyword
\code{EXT_MODEL} is set to the model of interest.  All suspected
extended sources are fitted with the model, allowing all of the
parameters to float.  The initial parameter guesses are critical here
to achieving convergence on the model fits in a reasonable time.  The
moments and the pixel flux distribution are used to make the initial
parameter guess.  Many of the source parameters can be accurately
guessed from the first and second moments.  The power-law slope can be
guessed by measuring the isophotal level at two elliptical radii and
comparing the ratio to that expected.

For each type of extended source model (in fact for all source
models), a function is defined which examines the fit results and
determines if the fit can be considered as a success or a failure.  The
exact criteria for this decision depends on the details of the model,
and so this level of abstraction is needed.  For example, in some
case, the range of valid values for each of the parameters must be
considered in the fit assessment.  In other cases, we may choose to
use only the parameter errors and the fit Chi-Square value.

All extended source model fits which are successful are then
subtracted from the image as is done for the successful PSF model
fits.  The background flux is retained, with the result that only the
source is subtracted from the image.  At this stage, the variance
image is not modified.  

For the single exposure (\ippstage{camera}) and \ippstage{stack} image
analysis, these galaxy model fits are only used internally to generate
a clean object-subtracted residual image.  For the PV3 analysis of the
$3\pi$ survey, these model fits were saved in the output catalog
files, but not loaded to the public database.  The \code{QGAUSS}
extended source model was used for the PV3 analysis (see
Section~\ref{sec:Source.Model}).  The convolved galaxy model fits (see
Section~\ref{sec:galaxy.conv.fit}) and the forced galaxy model fits
(see Section~\ref{sec:galaxy.forced.fit}) provide more reliable and
physically-motivated galaxy models.

For the difference image analysis, a trailed object model is used for
the extended sources; these model fit parameters are passed to the
Moving Object Processing System \citep[MOPS][]{2013PASP..125..357D}.

Any source which is fitted with the extended source model has the flag field
\code{PM_SOURCE_MODE_EXTMODEL} set.

\subsection{Faint Source Analysis}
\label{sec:faint.psf.model}

After a first pass through the image, in which the brighter sources
above a high threshold level have been detected, measured, and
subtracted, \ippprog{psphot} optionally begins a second pass at the image.  In
this stage, the new peaks are detected on the image with the bright
sources subtracted.  In this pass, the peak detection process uses the
variance image to test the validity of the individual peaks.  All peaks
with a significance greater than a user-defined minimum threshold are
accepted as sources of potential interest.  

The sources which are measured in this faint-source stage are clearly
low significance detections.  The PV3 threshold for the bright source
analysis is a signal-to-noise of 20.  The flag bit
\code{PM_SOURCE_MODE2_PASS1_SRC} is raised for sources detected in
this initial analysis stage.  The lower limit cutoff for the faint
source analysis in PV3 is a signal-to-noise of 5.0.  Sources detected
in the faint source stage are fitted with the PSF model using the
linear, ensemble fitting process.

In the \ippprog{psphotStack} version of the code, the 5 filter images
are processed together.  In this case, any source which is detected in
at least two of the five filters are then also measured on the other
filter images in which it was not detected above the signal-to-noise
limit.  The position in the other stack images is fixed based on the
pixel coordinates in the images in which the source was detected.
Detection in two filters is required in order to avoid excessive
forced photometry of spurious detections.  There is an interesting
class of astronomical objects which are extremely red (\eg, brown
dwarfs and high-redshift quasars).  Such sources are expected to be
detected only in the reddest filter (\yps).  For the $3\pi$ PV3
processing, we therefore also force the photometry in all filters for
sources which are only detected in \yps.  All sources which are forced
on one image based on detections in other images have the flag bit
\code{PM_SOURCE_MODE2_MATCHED} set.

\subsection{Extended Source Analysis}

After the initial, fast analysis of the image relying primarily on the
PSF model, a complete analysis of the extended source properties may
be performed.  For PS1 processing, this step is skipped in the nightly
(PV0) analysis of individual exposures and only performed for the
stacks in the major reprocessings.

The extended source analysis consists of the following types of
measurements: 1) an analysis of the radial profile of the surface
brightness of the source; 2) measurement of the Petrosian radius and
magnitude; 3) convolved galaxy model fits; and 4) photometry in
several fixed-sized apertures, both raw and convolved to a defined
PSF size.


The extended source analysis is not applied to all object which may be
galaxies.  Several restrictions are possible within the software.  For
example, it is possible to limit which objects are processed by their
apparent magnitudes, by their signal-to-noise, by an indication if they
are in fact extended, by the local stellar density, or by the galactic
latitude.  Some of these selections may be defined differently for the
galaxy model fits and the Petrosian parameters.

For the $3\pi$ PV3 processing, both an apparent magnitude cut and a
Galactic latitude cut were applied.  The apparent magnitude limits for
the galaxy model fits are applied to the measured Kron magnitude and
depend on the filter as follows: (\grizy) = (21.5, 21.5, 21.5, 20.5,
19.5).  These values were chosen to have roughly similar
signal-to-noise in a typical stack image for neutral color objects.
The magnitude limits for the Petrosian parameters were set to 25.0 for
all filters, far below the detection limits and effectively not
limiting the analysis based on apparent magnitude. For both galaxy
model fits and Petrosian parameters, the Galactic latitude cut was
defined by $|b| > b_{\rm min}$ where $b_{\rm min} = b_0 + r_b
e^{\frac{-l^2}{2 \sigma_b^2}}$.  For the PV3 analysis, $b_0 =
$20\degree, $r_b = $15\degree, $\sigma_b = $50\degree.  This contour
avoids the denser portions of the Galactic plane and bulge, limiting
the total time spent on the galaxy modeling analysis at the expense of
galaxy photometry in the plane (though Kron photometry is available
for those sources).  










\subsubsection{Radial Profiles}
\label{sec:radial.profile.v2}

Galaxies with regular profiles, such as elliptical galaxies and
regular spiral galaxies, may be described as primarily a radial
surface brightness profile, with additional structure acting as a
perturbation on that profile.  For many galaxies, the azimuthal shape
at a given isophotal level may be described as an elliptical contour.
To first order, a galaxy may be well described with a single elliptical
contour and radial profile.  

In order to facilitate the Petrosian photometry analysis below, \ippprog{psphot}
generates a radial profile for each suspected galaxy.  This analysis
starts by generating a radial profile in 24 azimuthal segments.  Near
the center of the galaxy, the profile is defined for radial
coordinates in steps of 1 pixel, with the closest pixel values
interpolated to that radial position.  Further from the center,
profile is defined using the median of the pixels landing in an
annular segment of size $\delta R = r \sin \theta$, rounded up to the
nearest integer pixel value.  The median of all pixels within a
rectangular approximation to the radial wedge is used.

The resulting 24 radial profiles are subject to contamination from
neighboring sources or to NAN values from masked pixels.  To clean the
profiles, pairs of radial profiles from opposite sides of the source
are compared.  Any masked values are replaced by the corresponding
value in the other profile.  The minimum of both profiles is then kept
for both profiles.  The result of this analysis is a set of profiles
of the form $f_i(r_i)$.  In this case, $f_i$ is effectively the
surface brightness for each radius in instrumental counts per pixel.
If fewer than 4 radial surface-brightness values are available for the
analysis, the source is skipped and the flag bit
\code{PM_SOURCE_MODE2_ECONTOUR_FEW_PTS} is set.  Some apparently
extended sources are in fact bright stars with central saturation.
These sources show up in this analysis as having many NAN-valued
pixels in the central regions.  During the radial profile analysis,
such sources are flagged with the bit
\code{PM_SOURCE_MODE2_RADBIN_NAN_CENTER} and are skipped from the rest
of the analysis.

The surface brightness profiles are then used to define the azimuthal
contour at a specific isophotal level.  This contour will be used to
rescale the radial profiles into a single set of profiles normalized
by the elliptical contour.  This contour is defined by determining the
median radius for profile bins with surface brightness in the range
$F_{\rm min} + 0.1 F_{\rm range}$ to $F_{\rm min} + 0.5 F_{\rm
  range}$.  The result of this analysis is a value for the radius as a
function of the angle for a well-defined surface brightness regime.
We then determine the elliptical shape parameters for this elliptical
contour: $R_{\rm major}, R_{\rm minor}, \theta$.  This ellipse is then
used to redefine a single radial profile normalized by the elliptical
contour: 
\[
\rho = \sqrt{\frac{x^2}{S^2_{xx}} + \frac{y^2}{S^2_{yy}} + x y S_{xy}} \\
\]

The surface brightness values are sampled at a number of radial
annuli, with the radii defined in the configuration
(\code{RADIAL.ANNULAR.BINS.LOWER} \&
\code{RADIAL.ANNULAR.BINS.UPPER}).  For each source, the resulting
surface brightness profile is saved in the output FITS table as a
vector (\code{PROF_SB}).  The flux at each radial position and the
fill-factor (fraction of pixels used to the total possible) are also
saved as equal-length vectors in the FITS table (\code{PROF_FLUX} and
\code{PROF_FILL}).  The values of the radial bins are saved in the
output file FITS header (\code{RMIN_NN}, \code{RMAX_NN}).  


\subsubsection{Petrosian Radii and Magnitudes}
\label{sec:petrosian}

\cite{1976ApJ...209L...1P} defined an adaptive aperture based on a
ratio of surface brightnesses.  The motivation is to define an
aperture which can be determined for galaxies without significant
biases as a function of distance from the observer.  Since surface
brightness in a resolved source is conserved as a function of
distance, using a ratio of surface brightness to define a spatial
scale results in a spatial scale which is constant regardless of
galaxy distance.

To measure the Petrosian radius and flux, we start by defining a
series of radial apertures with power-law spacing: $r_{i + 1} = 1.25
r_i$.  We calculate the surface brightness for the annulus from $r_i -
r_{i+1}$ by calculating the median of the values in the range $r_i /
\sqrt{1.25}$ to $r_{i+1} \sqrt{1.25}$ and dividing the the effective
area of the annulus corresponding to $r_i - r_{i+1}$.  

For any annulus $i$ spanning the radii $r_{\rm min}$ to $r_{\rm max} =
\beta r_{\rm min}$, the
Petrosian Ratio for that annulus is defined as the ratio of the
surface brightness in the annulus to the average surface brightness
within $r_{\rm max}$.  The Petrosian Radius is defined to be $r_{\rm
  max}$ for the annulus for which the Petrosian Ratio = 0.2, i.e., the
point on the galaxy radial profile at which the surface brightness is
20\% of the average surface brightness at that point.  If the profile
falls below the Petrosian ratio for the first radial bin, the flag bit
\code{PM_SOURCE_MODE2_PETRO_RATIO_ZEROBIN} is set to note that the
Petrosian radius may be poorly determined.

We determine the Petrosian Radius for the galaxy by quadratic
interpolation between the last two of the fixed annuli with Petrosian
Ratio $> 0.2$ and the first annulus with Petrosian Ratio $< 0.2$.  In
general, the Petrosian Ratio for a galaxy with a regular morphology
(spiral or elliptical) is falling monotonically, so this interpolation
is unambiguous.  However, irregular galaxy morphologies, noise, and/or
significant masking can cause the Petrosian Ratio to have rises as
well as drops.  We track the Petrosian Ratio until the value is no
longer significant ($\sigma_{\rm Ratio} < 2 {\rm Ratio}$).  If the
Petrosian ratio drops below 0.2 for more than one radius, we choose
the largest such radius.  If the Petrosian ratio does not fall below
0.2 for any of the measured radii, the annulus for which the ratio
falls to the lowest (yet still significant) value.  In such a case,
the flag bit \code{PM_SOURCE_MODE2_PETRO_INSIG_RATIO} is set.  

Once the Petrosian Radius has been determined, we can now measure the
Petrosian Flux : this is defined to be the total flux within an
aperture corresponding to 2 $\times$ the Petrosian Radius.  Using the
Petrosian Flux, we can calculate two other interesting radii: $R_{50}$
and $R_{90}$, the radii inside which 50\% and 90\% of the total
Petrosian flux is contained.  Sources for which the Petrosian
parameters are successfully measured have the flag bit
\code{PM_SOURCE_MODE_EXTENDED_STATS} set.  Sources for which the Petrosian
parameters were attempted, but for which the radial profile analysis
failed have the flag bit
\code{PM_SOURCE_MODE2_PETRO_NO_PROFILE} set.

\subsubsection{Convolved Galaxy Model Fits}
\label{sec:galaxy.conv.fit}

In the galaxy model fitting stage, sources which meet certain
criteria are fitted with analytical models for galaxies.  Three
traditional analytical galaxy models are implemented in \ippprog{psphot}
and used in the PV3 analysis:
\begin{itemize}
\item Exponential profile : $f = I_0 e^{-\rho}$
\item DeVaucouleur profile \citep{1948AnAp...11..247D}: $f = I_0 e^{-\rho^{1/4}}$
\item S\'ersic \citep{1963BAAA....6...41S} : $f = I_0 e^{-\rho^{1/n}}$
\end{itemize}
where $\rho$ is a normalized radial term: $\rho =
\sqrt{\frac{x^2}{R^2_{xx}} + \frac{y^2}{R^2_{yy}} + x y R_{xy}}$.  The
terms ($R_{xx}$, $R_{yy}$ , $R_{xy}$) describe the elliptical contour
and the profile scale in all three models and the coordinates $x$ \&
$y$ are determined relative to the centroids ($x,y = X_{\rm chip} -
x_0, Y_{\rm chip} - y_0$).  Including the normalization ($I_0$) and a
local sky value, the Exponential and DeVaucouleur profiles have 7 free
parameters and the S\'ersic profile has the additional free parameter of
the S\'ersic index $n$.  In this stage, the galaxy model is convolved
with an approximation to our best guess for the PSF model at the
location of the galaxy.

Sources which passed the extended source restrictions described above
were fitted with all three galaxy models, unless (a) the morphological
test identified the source as a likely cosmic ray
(Section~\ref{sec:source.size}) or (b) the peak of the PSF profile was
above the saturation limit for the chip \citep[see the discussion
  in][regarding the masking of saturated pixels]{waters2017}.  All
sources for which the extended source model fits were attempted the
flag bit \code{PM_SOURCE_MODE2_EXT_FITS_RUN} set.  If any of the
attempted model fits failed, then the flag bit
\code{PM_SOURCE_MODE2_EXT_FITS_FAIL} is set.  If all model fits
failed, then the flag bit \code{PM_SOURCE_MODE2_EXT_FITS_NONE} is
set.  


Before the non-linear fitting may be performed, it is necessary to
determine initial values for the parameters to be fitted.  For each of
the three model types, the position determined from the PSF fitting
analysis is used as the initial centroid $x_0,y_0$.  A guess for the
terms ($R_{xx}$, $R_{yy}$ , $R_{xy}$) is generated based on the second
moments.  The guess does not attempt to use the PSF model to adjust the
($R_{xx}$, $R_{yy}$ , $R_{xy}$) values; it was found that such a guess
tended to be too small and resulted in more iterations rather than
fewer. The 1st radial moment (see
\ref{sec:moments}) is used to estimate the effective radius of the
model based on the results of Graham \& Driver (2005, Table 1).  They
quantify the relationships between the first radial moment used to
calculated a Kron Magnitude and the effective radius for different
S\'ersic index values, $n$.  Since the Exponential and DeVaucouleur
models are equivalent to S\'ersic models with $n$ = 1 and 4,
respectively, this work can be used to generate the initial effective
radius values for all 3 model types.  Once the effective radius is
chosen, the second moments are used to define the aspect ratio and
position angle of the elliptical contour.  The Kron flux is used to
generate a guess for the normalization, applying an appropriate scale
factor based on the ($R_{xx}$, $R_{yy}$ , $R_{xy}$) values, generated
by integrating normalized S\'ersic models and determining the
relationship between the central intensity and the integrated flux as
a function of the S\'ersic index.

The PSF-convolved galaxy model fitting analysis uses the
Levenberg-Marquardt minimization method to determine the best fit.  In this
process, the $\chi^2$ value to be minimized is:
\[
\chi^2 (\bar{a}) = \sum_p \frac{1}{\sigma_p^2} \left[I_p - M_p(\bar{a}) \otimes \mbox{PSF} \right]^2 
\]
where $I_p$ represents the pixel values in the image (within some
aperture) and $M_p(\bar{a})$ represents the unconvolved galaxy model, a
function of a number of parameters $\bar{a}$, which is then convolved
with the PSF model.

We simplify this by defining:
\begin{eqnarray}
f_p (a_m)         & = & \frac{1}{\sigma_p} (I_p - M_p \otimes \mbox{PSF}) \\
\end{eqnarray}

To determine the minimization, we need the gradient and laplacian of
$\chi^2$ with respect to the model parameters, $a_m$:
\begin{eqnarray}
\chi^2 (\bar{a})  & = & \sum_p f_p^2  \\
2 \nabla   \chi^2  & = & \sum_p f_p \frac{\partial f_p}{\partial a_m} \\
\nabla^2 \chi^2  & \approx & H_{m,n} \\
2 H_{m,n}  & = & \sum_p \frac{\partial f_p}{\partial a_m} \frac{\partial f_p}{\partial a_n}
\end{eqnarray}
where we have approximated the Laplacian with the Hessian matrix,
$H_{m,n}$ by dropping the second-derivatives (which are assumed to be
a small perturbation).  Since
\[
\frac{\partial f_p}{\partial a_m} = -\frac{1}{\sigma_p}\frac{\partial M_p \otimes \mbox{PSF}}{\partial a_m}
\]
and since the order of the derivative and convolution may be
exchanged, we can write these in terms of the convolved image of the
model and the convolved images of the derivatives of the model $M_p$ with respect to the model parameters, $a_m$:
\begin{eqnarray}
\mathcal{M}_{p}   & = & M_p \otimes \mbox{PSF} \\
\mathcal{M}^\prime_{p,m} & = & \frac{\partial M_p}{\partial a_m} \otimes \mbox{PSF} \\
2 \nabla \chi^2    & = & -\sum_p \frac{I_p - \mathcal{M}_p}{\sigma_p} \mathcal{M}^\prime_{p,m} \\
2 H_{m,n}  & = &  \sum_p \frac{1}{\sigma_p^2} \mathcal{M}^\prime_{p,m} \mathcal{M}^\prime_{p,n}
\end{eqnarray}
The gradient vector and Hessian matrix are used in the
Levenberg-Marquardt minimization analysis using the standard
technique of determining a step from the current set of model
parameters to a new set by solving the matrix equation:
\[
(1 + \lambda_{m,n}) H_{m,n} = \delta \nabla \chi^2 
\]
where $\lambda_{m,n}$ is zero for $m \neq n$ and for $m = n$ set to be
large when the last iteration produced a large change in the
parameters compared to the local-linear expectation and small when the
last change was small.  The iteration ends when the change in the
parameters is small and/or the change in the $\chi^2$ value is small.

In the analysis, convolved galaxy fit, the galaxy model image and the
model derivative images must be convolved with the PSF at each
iteration step.  To save computation time, this convolution is
performed using a circularly symmetric approximation of the PSF model,
with the PSF model scale size set to the average of the major and
minor axis direction scale size of the full PSF model, with the same
radial profile term as the PSF model.  The convolution is performed
directly using the circular symmetry to reduce the number of
multiplications performed: all points in the 2D circularly symmetric
PSF model which have the same radial pixel coordinate can be evaluated
in the convolution by summing up the corresponding pixels in the
(galaxy model) image to be convolved before multiplying by the PSF
model profile at that radial coordinate.  This approximation reduces
the number of multiplications by a factor of \approx 8 for larger radii.
For the small size of the PSF model used to convolve the galaxy model
images, it was found that this direct convolution was faster than
using an FFT-based convolution.


For the Exponential and DeVaucouleur fits, all parameters are fitted
in the non-linear minimization stage.  For the S\'ersic model, we do not
fit the index within the Levenberg-Marquardt analysis.  Instead, we
start with a coarse grid search over a range of possible index values,
($n = 0.5, 1.0, 1.5, 2.0, 3.0, 4.0, 5.0, 6.0$) and a range of possible
values for $R_{\rm eff}$ based on the value of $R_1$, the first radial
moment.  For a given value of the S\'ersic index, the $R_{\rm eff}$ is
related to the 1st radial moment by the scale factor specified by
Graham \& Driver.  We use the observed value of the 1st radial moment
and try $R_{\rm eff}$ values of a factor of (0.8, 0.9, 1.0, 1.12,
1.25) times the value predicted by the Graham and Driver equation.
For each of these steps, the aspect ratio and position angle are held
constant and the normalization is determined to minimize the $\chi^2$.

We next perform 3 Levenberg-Marquardt minimization fits allowing the
shape parameters ($R_{xx}$, $R_{yy}$ , $R_{xy}$) and the normalization
to be fitted, holding the centroid ($x_0, y_0$), S\'ersic index $n$,
and sky constant.  In these fits, the index $n$ is set to the minimum
value previously calculated as well as values halfway to the next, and
previous, values in the grid above.  E.g., if the minimum fitted index
value is 3.0, then the LMM fits are performed using $n$ = 2.5, 3.0,
3.5.  The resulting $\chi^2$ values are then used to perform quadratic
interpolation to find the index $n$ which produces the locally minimum
$\chi^2$ value.  Finally, this best-fit index value is held constant
while Levenberg-Marquardt minimization is used to find the best fit
values of all other parameters.  Sources for which a convolved galaxy
model fit was successful have the flag bit
\code{PM_SOURCE_MODE_EXTENDED_FIT} set.


The central pixel of the S\'ersic, DeVaucouleur, and Exponential
models requires special handling.  When comparing an analytical model
to the pixelized image recorded by a CCD, one normally treats the
value in a pixel as equivalent to the value of the model at the center
of the pixel.  However, in reality, the number of counts observed in a
pixel represents the integral of the surface brightness across the
area of the pixel.  This average will be equal to the central surface
brightness times the area of a pixel as long as the second and higher
derivatives of the analytical model are zero.  However, if the first
and second derivatives are non-zero, the curvature of the function
within the pixel will make the integral differ from the central
surface brightness times a fixed pixel area.  If the curvature of the
model function is sufficiently large, this difference will have a
significant impact on the evaluation of the model.   This situation is
particularly true for the central portion of the S\'ersic-like model
functions. 


In order to accurately compare the observed galaxy flux distribution
to a model, it is necessary to integrate the pixel flux for a given
set of model parameter values.  This could be done in a numerical
fashion, but in practice brute-force evaluation of the numerical
integral is computationally very expensive, requiring many evaluations
of the model function.  Within \ippprog{psphot}, we bypass this
problem by defining a set of pre-calculated images for the central 9
pixels (the $3 \times 3$ grid of pixels centered on the peak).  These
pixel images are defined at higher resolution, with 11 subpixels per
real CCD pixel.  The pre-calculated images are generated for a series
of values for the following parameters: S\'ersic index, effective
radius, axial ratio.  We then select the closest image to our specific
case, and integrate over the true sub-pixels relevant for our position
and model.  We have thus turned the problem from thousands of
evaluations of the full galaxy model to \approx 100 straight
additions, or up to $6 \times$ that number if we interpolate between
any of the parameters.

\subsubsection{Fixed Aperture Photometry}
\label{sec:fixed.aperture.photom}

For some science goals, a well-measured color of a galaxy is more
important than an accurate total magnitude.  In the case of PS1, the image
quality variations for stacks of different filters presents a serious
challenge for the determination of precise colors.  \ippprog{psphot} determines
a set of PSF-matched radial aperture flux measurements in order to
minimize the impact of the stack image quality variations.

In \ippprog{psphotStack}, the stack analysis version of \ippprog{psphot},
the 5 filter images are processed together.  After the PSF models have
been fitted and a best set of galaxy models have been determined,
three sets of fixed circular apertures are measured.  In the first
set, the fluxes in the apertures are measured using the raw stack
images.  The centers of the apertures for each source across the 5
filters are fixed so that the pixels represent the equivalent portions
of the same galaxy for all 5 filters.  In this analysis, the best
model for each source is subtracted from the image pixels for all
sources excluding the source in consideration.  The 'best model' is
determined based on the minimum $\chi^2$ value for the model fits.

In addition to the raw fixed circular apertures, the stack images are
each convolved with a circular Gaussian with $\sigma$ chosen to yield
an image with a typical FWHM of 6 pixels (1.5\arcsec).  The full set
of circular apertures are again measured on these convolved images.
Again, the best source models are subtracted from the image for
sources not being measured.  This subtraction includes the convolution
to smooth the model to the effective FWHM of the convolved image.  The
entire procedure is then repeated with a target FWHM of 8 pixels
(2\arcsec).

For the PV3 analysis of the $3\pi$ survey data, the fluxes are
measured for a set of up to 9 circular apertures with sizes chosen to
match the similar circular apertures measured by the SDSS analysis.
These apertures have radii of (4.16, 7.04, 12.0, 18.56, 29.76, 45.68,
72.80, 112.80, 176.88) pixels = (1.04, 1.76, 3.00, 4.64, 7.44, 11.42,
18.20, 28.20, 44.22) arcseconds.  If the object is too faint, the
larger apertures will be largely noise and the computation is
wasteful.  We only calculate the circular apertures out to the second
aperture larger than the ``sky radius'' (defined in
Section~\ref{sec:radial.profile}), but we calculate photometry for at
least the smallest 4 apertures.  Sources for which photometry in these
fixed aperture are calculated have the flag bit
\code{PM_SOURCE_MODE_RADIAL_FLUX} set.


\subsection{Aperture Correction and Total Aperture Fluxes}
\label{sec:aperture.correction}

A PSF model will always fail to describe the flux of the stellar
sources at some level.  For high-precision photometry, we need to be
able to correct for the difference between the PSF model fluxes and
the total flux of the sources.  In the end, all astronomical
photometry is in some sense a relative measurement between two images.
Whether the goal is calibration of a science image taken at one
location to a standard star image at another location, or the goal is
simply the repetitive photometry of the same star at the same location
in the image, it is always necessary to compare the photometry between
two images.  If this measurement is to be consistent, then the
measurement must represent the flux of the stars in the same way
regardless of the conditions under which the images were taken, at
least within some range of normal image conditions.  So, for example,
two images with different image quality, or with different tracking
and focus errors, will have different PSF models.  To the extent the
PSF model is inaccurate, the measured flux of the same source in the
two images will be different (even assuming all other atmospheric and
instrumental effects have been corrected).  The amplitude of the error
will by determined by how inconsistently the models represent the
actual source flux.

Aperture photometry attempts to avoid these problems, but introduces
other difficulties.  In aperture photometry, if a large enough
aperture is chosen, the amount of flux which is lost will be a small
fraction of the total source flux.  Even more importantly, as the
image conditions change, the amount lost will change by an even
smaller fraction, at least for a large aperture.  This can be seen by
the fact that the dominant variations in the image quality are in the
focus, tracking and seeing.  All of these errors initially affect the
cores of the stellar images, rather than the wide wings.  The wide
wings are largely dominated by scattering in the optics and scattering
in the atmosphere.  The amplitude and distribution of these two
scattering functions do not change significantly or quickly for a
single telescope and site.  Aperture photometry can then be used to
correct the PSF photometry.

The difficulty for aperture photometry is the need to make an accurate
measurement of the local background for each source.  As the aperture
grows, errors in the measurement of the sky flux start to become
dominant.  If the aperture is too small, then variations in the image
quality are dominant.  The brighter is the source, the smaller is the
error introduced by the large size of the aperture.  However, the
number of very bright stars is limited in any image, and of course the
brighter stars are more likely to suffer from non-linearity or
saturation.  

In order to thread the needle between these effects, \ippprog{psphot}
measures the aperture photometry on a modest-sized aperture, and then
uses the PSF model to extrapolate to a large aperture.  When the PSF
fluxes are calculated, the aperture flux for the modest-sized aperture
is also determined.  The aperture is a circular aperture with radius
set to a fixed multiple (\code{PSF_APERTURE_SCALE}) of $\sigma_w$, the
width of the Gaussian window function determined based on the analysis
of the second moments (see Section~\ref{sec:moments}).  For the PV3
$3\pi$ analysis, the aperture window radius is $4.5 \times \sigma_w$,
while the large reference aperture radius is set to 25 pixels
(\code{PSF_REF_RADIUS} = 6\farcs4).  These corrected aperture
magnitudes are saved in the output catalogs as \code{AP_MAG}, the
uncorrected aperture magnitudes are saved as \code{AP_MAG_RAW}, and
the radius used to measure the raw aperture flux is saved as
\code{AP_MAG_RADIUS}.  The corresponding flux and the flux error are
saved as \code{AP_FLUX} and \code{AP_FLUX_SIG}.

With these aperture magnitudes in hand, it is now possible to make an
average correction to the PSF magnitudes to bring the PSF and aperture
magnitudes to the same system.  This correction is measured using the
same stars from which the PSF model is measured, as long as the PSF
magnitude error for the star is less than 0.03 mag.  The correction is
calculated using the weighted average of the values $m_{\rm AP} -
m_{\rm PSF}$.  Since the PSF may vary across the image, the correction
is determined as a function of position in the image.  Like the PSF
model, the 2D variations of the aperture correction may be modeled as
a polynomial or via interpolation in a grid.  For the $3\pi$ PV3
analysis, a grid with a maximum of $6\times 6$ samples per GPC1 chip
image was used.  The reported PSF magnitudes for all objects have this
aperture correction applied.

\ippprog{psphot} allows a collection of PSF model functions to be tried on all
PSF candidate sources.  For each model test, the above corrected
ApResid scatter is measured.  The PSF model function with the smallest
value for the ApResid scatter is then used by \ippprog{psphot} as the best PSF
model for this image.  The number of models to be tested is specified
by the configuration keyword \code{PSF_MODEL_N}.  The configuration
variables \code{PSF_MODEL_0}, \code{PSF_MODEL_1}, through
\code{PSF_MODEL_N - 1} specify the names of the models which should be
tested.

\section{Forced Photometry Modes}
\label{sec:psf.forced.fit}

Traditionally, projects which use multiple exposures to increase the
depth and sensitivity of the observations have generated something
equivalent to the stack images produced by the IPP analysis
(c.f, CFHT Legacy survey, COSMOS, etc).  In theory, the photometry of
the stack images produces the ``best'' photometry catalog,
with best sensitivity and the best data quality at all magnitudes.  In
practice, these images have some significant limitations due to the
difficulty of modeling the PSF variations.  This difficulty is
particularly severe for the Pan-STARRS $3\pi$ survey stacks due to the
combination of the substantial mask fraction of the individual input
exposures, the large intrinsic image quality variations within a
single exposure, and the wide range of image quality conditions under
which data were obtained and used to generate the $3\pi$ PV3 stacks.

For any specific stack, the point spread function at a particular
location is the result of the combination of the point spread
functions for those individual exposures which went into the stack at
that point.  Because of the high mask fraction, the exposures which
contributed to pixels at one location may be somewhat different just a
few tens of pixels away.  In the end, the stack images have
a effective point spread function which is not just variable, but
changing significantly on small scales in a highly textured fashion.

Any measurement which relies on a good knowledge of the PSF at the
location of an object needs to determine the PSF variations present in
the stack image or the measurement will be somewhat degraded.  The
highly textured PSF variations make this a very challenging problem:
not only would such a PSF model require an unusually fine-grained PSF
model, there would likely not be enough PSF stars in a given stack
image to determine the model at the resolution required.  The IPP
photometry analysis code uses a PSF model with 2D variations using a
grid of at most $6\times 6$ samples per skycell, a number reasonably
well-matched to the density of stars at most moderate Galactic
latitudes.  This scale is far too large to track the fine-grained
changes apparent in the stack images.

As a result, PSF photometry as well as convolved galaxy models in the
stack are degraded by the PSF variations.  Aperture-like measurements
are in general not as affected by the PSF variations, as long as the
aperture in question is large compared to the FWHM of the PSF.


The IPP analysis solves this problem by starting with the sources
detected in the stack images and performing forced photometry on the
individual warp images used to generate the stack, and then combining
the resulting measurements to determine a high-quality average value.
This forced-photometry analysis is performed using the
\ippprog{psphotFullForce} variant of \ippprog{psphot}.

In this program, the positions of sources are loaded from the output
catalog of the stack photometry.  Candidates PSF stars are
pre-identified as those stars used to generate the PSF model in the
stack photometry analysis.  A PSF model is generated for each input
warp image based on those stars; PSF stars which are excessively
masked on a particular image are not used to model the PSF.  The PSF
model is fitted to all of the known source positions in the warp
images.  Aperture magnitudes, Kron magnitudes, and moments are also
measured at this stage for each warp.  Note that the flux measurement
for a faint, but significant, source from the stack image may be at a
low significance (less than the $5\sigma$ criterion used when the
photometry is not run in this forced mode) in any individual warp
image; the measured flux may even be negative due to statistical
fluctuations.  When combined together, these low-significance
measurements result in a significant measurement as the signal-to-noise
increases with the combination of more data.

Individual warp images are processed independently with separate
executions of the \ippprog{psphotFullForce} program.  Sources which
are loaded by \ippprog{psphotFullForce} for analysis are marked with
the flag bit \code{PM_SOURCE_MODE_EXTERNAL}.  This bit is also used to
mark user-supplied sources loaded for analysis by the regular version
of \ippprog{psphot}.  Once all of the forced photometry measurements
(for point sources as well as galaxies, discussed below) are completed
for all of the warps which contributed to a stack image, the
measurements are combined together by other portions of the IPP
system.  The PSF photometry measurements are combined in the context
of the DVO database system \citep{magnier2017.datasystem}, including
recalibration of the zero points for the individual warp.

\subsection{Forced Galaxy Models}
\label{sec:galaxy.forced.fit}

The convolved galaxy models are also re-measured on the warp images by
the \ippprog{psphotFullForce} analysis.  In this analysis, the galaxy
models determined from the stack image analysis are used to seed the
analysis in the individual warp images.  The motivation of this
analysis is the same as the forced PSF photometry: the PSF of the
stack image is poorly determined due to the masking and PSF variations
in the inputs.  Without a good PSF model, the PSF-convolved galaxy
models are of limited accuracy.

In the forced galaxy model analysis, we assume that the galaxy
position and position angle, along with the S\'ersic index if
appropriate, have been sufficiently well determined in the analysis of
the stack image.  In this case, the goal is to determine the best
values for the major and minor axis of the elliptical contour and at
the same time the best normalization corresponding to the best
elliptical shape, and thus the best galaxy magnitude value.

For each warp image, the stack values for the major and minor axis are
used as the center of a grid search of the major and minor axis
parameter values.  The grid spacing is defined as a function of the
signal-to-noise of the galaxy in the stack image so that bright
galaxies are measured with a much finer grid spacing than faint
galaxies.  For the PV3 $3\pi$ analysis, a $5 \times 5$ grid was used;
values in both the major and minor axis directions of ($1 -
\frac{3.0}{S/N}$, $1 - \frac{1.5}{S/N}$, 1.0, $1 + \frac{1.5}{S/N}$,
$1 + \frac{3.0}{S/N}$) times the dimension are tested.  For each grid
point, the major and minor axis values at that point are used to
generate the model.  The model is then convolved with the PSF model
for the warp image at that point.  The resulting convolved model is
then compared to the warp pixel data values and the best fit
normalization value is determined.  The integrated flux, flux error,
and the $\chi^2$ value for each grid point are recorded.

For a given galaxy, the result is a collection of $\chi^2$ values,
fluxes, and flux errors for each of the grid points spanning all
warp images.  A single $\chi^2$ grid can then be made by
combining each grid point across the inputs.  The combined $\chi^2$
for a single grid point is simply the sum of all $\chi^2$ values at
that point.  If, for a single warp image, the galaxy model
is excessively masked, then that image will be dropped for all grid
points for that galaxy.  The reduced $\chi^2$ values can be determined
by tracking the total number of pixels used across all inputs to
generate the combined $\chi^2$ values.  From the combined grid of
$\chi^2$ values, the point in the grid with the minimum $\chi^2$ is
found.  Quadratic interpolation is used to determine the major, minor
axis values for the interpolated minimum $\chi^2$ value.  The errors
on these two parameters is then found by determining the contour at
which the $\chi^2$ increases by 1.

In this way, the forced galaxy model analysis uses the PSF information
from each warp image to determine a best set of convolved galaxy
models for each galaxy model measured for the stack image.

\section{Difference Image Photometry}

Among the primary science drivers for Pan-STARRS are the detection of
moving objects (e.g., asteroids) and explosive transient sources
(e.g., supernovae).  For both of these situations, difference images
are commonly used to remove the clutter of the static stars and
galaxies.  In the Pan-STARRS system, difference images are generated
using the PSF-matching technique described by
\citep[e.g.,][]{1998ApJ...503..325A}.  The description of the
Pan-STARRS implementation is given by \cite{price2017}.  The analysis
of the sources detected in these difference images uses a portion of
the \ippprog{psphot} code embedded in the program, \ippprog{ppSub},
which generates those image.  

The analysis of the difference image follows the same basic steps as
other \ippprog{psphot} versions with some minor modifications (see
Table~\ref{tab:measurements}), as follows.  The background subtraction
is performed before the PSF matching and image subtraction is
performed.  The PSF model construction stage is not possible in the
difference image due to the lack of valid sources.  Instead, the PSF
model from is generated from the positive image, after PSF-matching
but before the subtraction is performed.  Because we do not expect to
have a large number of sources, only a single source detection pass is
performed, and at the lowest signal-to-noise threshold.  Only linear
PSF model fitting is performed using the centroid determined from the
analysis of the source moments.  

For the difference images, the galaxy model analysis is not relevant.
In a properly-constructed difference image, galaxies are unlikely to
remain behind as significant sources.  Most real sources in the
difference image will be PSF-like and will consist of photometrically
variable sources (flare stars, supernovae, etc) or astrometrically
variable sources (high-proper motion stars or solar-system bodies).
There are three likely classes of sources which will not be well
represented by the PSF model, as discussed below.

Fast-moving solar-system objects will appear as short streaks.  For
example, a fast solar system object may have an apparent rate of 0.5
degrees per hour, translating to 15 arcseconds in a 30 second
exposure.  Even a main belt asteroid at roughly 1 AU has reflex motion
of approximately 1 degree per day, equivalent to 1.25 arcsec in a 30
second exposure, and may be noticeably smeared and non-PSF-like.  In
\ippprog{psphot}, we use a trailed-star model to characterize these
types of sources.  This model is fitted in the same portion of the
code which performs the unconvolved galaxy model analysis.

In some cases, the stars in the two images may be somewhat offset.
For specific stars, this offset may be due to differential chromatic
aberration from the atmosphere or the optics, or from modest proper
motion.  If the astrometric solution for one of the two images is
insufficiently accurate, all stars in large portions of the images may
be noticeably displaced.  In both of these situations, the stars will
appear as PSF dipoles in the difference images.  The positive and the
negative images will have stellar profiles, but they will be offset
and will not subtract well.  The two components may not have the same
amplitude.  In theory, a PSF-dipole model could be used to fit these
types of sources, with free parameters of the two centroids and the
two fluxes.  In practice in \ippprog{psphot}, we use a number of
non-parametric pixel-level statistics in an attempt to detect these
cases.

For the difference images, we measure the following quantities for
each of the detections, using only pixels within the photometry
aperture.  First, we count the number of masked pixels (\code{nMask}),
the number of pixels with positive flux (\code{nGood}), and the number
of pixels with negative flux (\code{nBad}).  We also add the total
flux in positive pixels (\code{fGood}) and total absolute value of the
flux in negative pixels (\code{fBad}).  Using these values, We report
the following quantities:
\begin{itemize}
\item \code{nGood}
\item \code{fRatio} = \code{fGood} / (\code{fGood} + \code{fBad})
\item \code{nRatioBad} = \code{nGood} / (\code{nGood} + \code{nBad})
\item \code{nRatioMask} = \code{nGood} / (\code{nGood} + \code{nMask})
\item \code{nRatioAll} = \code{nGood} / (\code{nGood} + \code{nMask} + \code{nBad})
\end{itemize}

We also attempt to place the difference image detections in the
context of the input images, both the positive (subtrahend) and
negative (minuend) images.  We identify the closest source in both the
positive and negative images to the detection in the difference image,
out to a maximum of \code{INPUT.MATCH.RADIUS} (= 50 pixels), but only
if the source in those images has a signal-to-noise greater than
\code{INPUT.MATCH.MIN.SN} (= 10).  If there is a close neighbor in the
positive image, and the difference in the magnitudes of the source in
that image and the source in the difference image is less than 5
$\sigma$, then the bit \code{PM_SOURCE_MODE2_DIFF_SELF_MATCH =
  0x00000800} is raised in \code{mask2} as these two detections are
likely the same flux (\ie, detection of an isolated source).  

If the difference image detection is matched to a nearby source in the
positive image, then the signal-to-noise of the neighbor is saved as
\code{DIFF_SN_P} and the distance in pixels between the difference
detection and positive detection is saved as \code{DIFF_R_P}.
Similarly, for a neighbor in the negative image, these values are
saved as \code{DIFF_SN_M} and \code{DIFF_R_M}.  Additional
\code{mask2} bits are also raised: if the difference detection is only
associated with one of the two input images, then the bit
\code{PM_SOURCE_MODE2_DIFF_WITH_SINGLE = 0x00000001} is raised, while
a difference detection which has a match in both input images has
\code{PM_SOURCE_MODE2_DIFF_WITH_DOUBLE = 0x00000002} raised. 

Comets appear in the difference images as a non-PSF sources.  Their
2-D structure includes both the flux from the coma (with a typical
power-law profile) and flux from the tail (with a more complex flux
distribution).  We use the Kron magnitudes to identify possibly
extended objects which may be cometary in nature.


For a difference image, both positive and negative sources will be
present.  The basic peak detection algorithm will only trigger for the
positive sources.  In the \ippprog{ppSub} program, both the $A - B$
and the $B - A$ images are sent to the \ippprog{psphot} routine for
source detection and characterization.

Note that the variance image for a difference image must be generated
from the two positive images used to construct the difference.  It is
possible to run \ippprog{psphot} as an external program on a
difference image generated previously.  In this case, the variance
image and the PSF model must be supplied as well as the difference
image.



\acknowledgments

The Pan-STARRS1 Surveys (PS1) have been made possible through
contributions of the Institute for Astronomy, the University of
Hawaii, the Pan-STARRS Project Office, the Max-Planck Society and its
participating institutes, the Max Planck Institute for Astronomy,
Heidelberg and the Max Planck Institute for Extraterrestrial Physics,
Garching, The Johns Hopkins University, Durham University, the
University of Edinburgh, Queen's University Belfast, the
Harvard-Smithsonian Center for Astrophysics, the Las Cumbres
Observatory Global Telescope Network Incorporated, the National
Central University of Taiwan, the Space Telescope Science Institute,
the National Aeronautics and Space Administration under Grant
No. NNX08AR22G issued through the Planetary Science Division of the
NASA Science Mission Directorate, the National Science Foundation
under Grant No. AST-1238877, the University of Maryland, and Eotvos
Lorand University (ELTE) and the Los Alamos National Laboratory.

\bibliographystyle{apj}

\end{document}